\documentclass[epj]{svjour}
\usepackage{amsmath,graphicx,slashed,indentfirst,fixltx2e}
\usepackage{multirow}

\begin{document}
\title{Finite-width Gaussian sum rules for $0^{-+}$ pseudoscalar glueball based on correction from instanton-gluon interference to correlation function}
\author{Feng Wang \and Junlong Chen \and Jueping Liu\inst{}
\thanks{Author to whom all correspondences should be addressed.
E-mail address: jpliu@whu.edu.cn}%
}                     
%
%
\institute{Department of Physics,School of Physics Science and Technology,Wuhan University,430072,Wuhan,China}
\date{Received: date / Revised version: date}
%
\abstract{Based on correction from instanton-gluon interference to correlation function, the properties of the $0^{-+}$ pseudoscalar glueball is investigated in a family of finite-width Gaussian sum rules. In the framework of semiclassical expansion for quantum chromodynamics $(\textrm{QCD})$ in the instanton liquid background, the contribution arising from the the interference between instantons and the quantum gluon fields is calculated, and included in the correlation function together with  pure-classical contribution from instantons and the perturbative one. The interference contribution is turned to be gauge-invariant, free of infrared divergence, and has a great role to restore the positivity of the spectra of the full correlation function. The negligible contribution from vacuum condensates is excluded in our correlation function to avoid the double counting. Instead of the usual zero-width approximation for the resonances, the usual Breit-Wigner form with a suitable threshold behavior for the spectral function of the finite-width resonances is adopted. A consistency between the subtracted and unsubtracted sum rules is very well justified. The values of the mass, decay width and coupling constants for the $0^{-+}$ resonance in which the glueball fraction is dominant are obtained, and agree with the phenomenological analysis.
} 
\maketitle
%

\section{Introduction}
A significant issue in quantum chromodynamics (QCD) is to seek for the signal of the existence of glueballs. Because glueballs are bound sates composed of only gluons in the quarkless world, such signal may give a unique insight into the non-Abelian dynamics of QCD. Theoretical investigations including lattice simulations \cite{Phys.Rev.Lett.57.1288,Phys.Rev.D60.114501,Phys.Rev.D80.114502},  model researches \cite{Phys.Rev.Lett.54.869,Phys.Lett.B395.123,Phys.Rev.Lett.101.252002} and sum rule analyses \cite{PhysRevD.71.054008,J.Phys.G19.373,PhysRevLett.75.1707,Phys.Lett.B523.127,NuclPhysA.695.205,PhysRevD.79.014024} have been going on for a long time, but no decisive evidence of the existence of glueballs has been confirmed by experimental research up to now \cite{J.Phys.G32.R293,Prog.Part.Nucl.Phys.63.74}. Further investigation on glueballs still makes sense.

One of the obstacles in theoretical researches of glueballs is that non-perturbative dynamics of QCD, which is responsible for the formation of hadrons, is difficult to handle, and the QCD vacuum is recognized to be a medium with complicated structure, and may impact greatly on the attributes of hadrons. In particular, the tunneling effect between the degenerate vacua of QCD should be taken into account. In the leading order, this effect is
described by instantons \cite{PhysRevD.14.3432,RevModPhys.70.323} and shown to be of great significance in generating the properties of the unusual hadrons, glueballs. Moreover, the glueball may be mixed with usual mesons of the same quantum numbers, making the identification of the glueball more complicated \cite{PhysRevD.79.014024,PhysRevD.73.054006}.

Instantons, as the strong topological fluctuations of gluon fields in QCD, are widely believed to play an important role in the physics of the strong interaction (for reviews see \cite{RevModPhys.70.323,ProgPartNuclPhys.51.173}). In particular, instantons provide mechanisms for the violation of both $U(1)_A$ and chiral symmetry in QCD, and may therefore be important in determining hadron masses
and in the resolution of the famous $U(1)_A$ problem. Furthermore, it was recently shown that instantons persist through the deconfinement transition, so that instanton-induced interactions between quarks and gluons may underlie the unusual properties of the so called strongly coupled quark-gluon plasma recently discovered at RHIC \cite{Shuryak:2006se}.



In the instanton liquid model, a narrow sense describes the QCD vacuum as a sum of independent instantons with radius $\bar{\rho} = (600\mathrm{MeV})^{-1}$ and
effective density $\bar{n} = (200\textrm{MeV})^4$ \cite{Nucl.Phys.B184.443}. This model avoid the infrared problem caused by an infinite instanton density in the diluted gas model. The correctness of the instanton liquid model is still being intensively investigated. So far the model is essentially justified by its phenomenological success. The most important predictions are probably the breaking of the chiral symmetry (SBCS) in the axial triplet channel \cite{Nucl.Phys.B.245.259,Nucl.Phys.B.272.457} and the absence of Goldstone bosons in the axial singlet channel.

The non-perturbative effects of QCD is commonly casted into the description of non-perturbative vacuum, such as quark and gluon condensates and the instanton configurations. When contributions from both condensates and the instanton are included into the correlation function together, it leads to a double counting in a
sense that they can be two alternative ways for parametrizing the nonperturbative vacuum
\cite{Narison:2002hk,Phys.Lett.B522.266}. Moreover, the instanton distribution is closely connected with the vacuum condensates since the mean size and density of instantons can be deduced from the quark and gluon condensates, and conversely, the values of condensates can be reproduced from instanton distribution \cite{NuclPhysB.203.116,Phys.Lett.B147.351,PhysRevLett.49.259,J.Phys.G21.751}. The contributions of instanton and those of the condensates should be equivalent from each other in some case, and thus including both contributions at the same time will cause the double counting problem. To avoid it, many authors invoke some special techniques in dealing with the two contributions \cite{AnnalsPhys.254.328,PhysRevD.64.034015,Phys.Rev.D85.054007,Phys.Rev.D78.076005}. This issue is, however, really not settled. Fortunately, the contribution of condensates to the correlation function in the glueball channels is unusually weak as demonstrated in this work and by many other authors\cite{Phys.Lett.B86.347,Nucl.Phys.B509.312}. Therefore, it is assumed that the condensate contributions can be understood as a small fraction of the corresponding instanton one in the local limit. To prevent thoroughly from the problem of double counting, we choose to work in the instanton vacuum model of QCD, and carry out a semi-classical expansion in instanton background fields as suggested in our previous works to analyze the properties of the lowest $0^{++}$ scalar glueball \cite{ChinPhysLett.23.2920,PhysRevD.82.016003} and $0^{-+}$ pseudoscalar one \cite{Eur.Phys.J.Plus128.115,J.Phys.G.41.035004}, where the correlation function of the glueball currents are calculated by just including the contributions from the pure instantons, the pure quantum gluons, and the interference between both, instead of working with both instantons and condensates at the same time.

Zhang and Steele \cite{NuclPhysA.728.165} have reported a disparity between their Laplace sum rule and Gaussian sum rule for the pseudoscalar glueball. In this reference, the optimized parameters of pseudoscalar glueball have been obtained from the Gaussian sum rule, while it fails to achieve a satisfying Laplace sum rule. This result seems strange, since both Laplace and Gaussian sum rules are derived from the same underlying dynamical theory, and should, at least, be approximately consistent. It reflects the inconsistency for including both condensate and instanton contributions in the correlation function and disregarding the important contribution from the interaction between instantons and their quantum counterparts at the same time.

An other serious problem in the $0^{-+}$ glueball sum rule approach is that the fundamental spectral positivity bound is violated when including the strong repulsive pure instanton contribution, and as a consequence, the signal for the pseudoscalar glueball disappears\cite{NuclPhysA.728.165}. To cure this pathology of positivity violation, the topological charge screening effect in the QCD vacuum is added to the correlation function, and a suitable instanton size distribution is taken into account\cite{Braz.J.Phys.34.875,PhysRevD.71.054008}. However, as comparing with the interference contribution, which we have recalculated in this paper, and the pure perturbative one in the considered energy region, the topological screening effect turns out to be negligible; and the pathology of positivity violation disappears when including the interference contribution (s. below).

Phenomenologically, the identification of the pseudoscalar glueball has been a matter of debate since the Mark II experiment proposed glueball candidates \cite{PhysLettB.97.329}. Later, in the mass region of the first radial excitation of the $\eta$ and $\eta'$ mesons, a supernumerous candidate, the $\eta(1405)$ has been observed.
It turns out to be  clear that $\eta(1405)$ is allowed as glueball dominated state mixed with isoscalar $q\bar{q}$ states due to its behavior in production and decays, namely, it has comparably large branching ratios in the $J/\psi$ radiative decay, but not been observed in $\gamma \gamma$ collisions\cite{Int.J.Mod.Phys.E18.1,J.Phys.G32.R293,PhysLettB.501.1}. A review on the experimental status of the $\eta(1405)$ is given in Ref. \cite{J.Phys.G32.R293}. However, this state lies considerably lower than the theoretical expectations: the lattice QCD predictions suggest a glueball around $2.5$GeV \cite{PhysRevD.58.055003,PhysRevD.60.034509}; the mass scale of the pseudoscalar glueball obtained in the QCD sum rule approach is above $2$GeV \cite{Eur.Phys.J.Plus128.115,J.Phys.G.41.035004,Braz.J.Phys.34.875,PhysRevD.71.054008}. On the other hand, there are attractive arguments for the approximately degenerate in mass for the scalar and pseudoscalar glueballs \cite{PhysRevD.70.114033}, and even the scenario that a pseudoscalar glueball may be lower in mass than the scalar one is recently discussed in Ref. \cite{Phys.Rev.D81.014003}.
The possibly non-vanishing gluonium content of the ground state $\eta$ and $\eta'$ mesons is discussed in \cite{JHEP.0710.026,Phys.Lett.B648.267,JHEP.0705.006,PhysRevD.79.014024}.
Up to now, only a topological model of the glueball as closed flux tube \cite{PhysRevD.70.114033} predicts degeneracy of the $0^{++}$ and $0^{-+}$ glueball masses and admits the region
$1.3$-$1.5$GeV.

To face a more realistic phenomenological situation, we now reexamine the correlation function in the instanton liquid vacuum, and include all the resonances below and near the $\eta(1405)$ into the finite-width spectral function, and then achieve a series of results in traditional Gaussian sum rule analyses which are consistent with the phenomenology. On the other hand, as a crosscheck, these results are almost same with the Laplace ones (paper is submitted), because both Laplace and Gaussian sum rules are derived from the same underlying dynamical theory. The paper is organized as follows:  In the second section, we present systematically the calculation of the contribution to the correlation function due to the interference between instantons and quantum gluons. It is this interference between instantons and quantum gluons to serve to be a mechanism to keep the positivity of the spectral function in contrast with the so-called topological charge screening effect stressed in Ref.\cite{PhysRevD.71.054008}. The effect of the contribution of topological charge screening is found to be negligible as comparing with the interference one in the considered energy region. The spectral function is constructed in the similar way as in case of the $0^{++}$ glueball\cite{PhysRevD.82.016003,JPhysG.38.015005} with a suitable threshold behavior in the third section. In the fourth section, a family of Gaussian sum rules are constructed. The numerical simulations are carried out in the fifth section, and the results are consistent with various Gaussian sum rules and in accordance with the phenomenology. Finally, the main conclusions are given, and a discussion of some interesting issues is open.

\section{Correlation function}
We are working in Euclidean QCD. The pseudoscalar glueball current is defined as
\begin{eqnarray}
&O_{p}(x)&=\alpha_{s}G^{a}_{\mu\nu}(x)\widetilde{G}_{\mu\nu,a}(x)
\label{eq2:1}
\end{eqnarray}
where $\alpha_s$ is the strong coupling constant, $G^{a}_{\mu\nu}(x)$ is the gluon field strength tensor with the color index $a$ and Lorentz indices $\mu$ and $\nu$, and
\begin{eqnarray}
\widetilde{G}_{\mu\nu,a}(x)=\frac{1}{2}\epsilon_{\mu\nu\rho\sigma}G^{a}_{\rho\sigma}(x)
\label{eq2:2}
\end{eqnarray}
is the dual of $G^{a}_{\mu\nu}(x)$. The current $O_p(x)$ is a Lorentz-irreducible, gauge-invariant and local composite operator with the lowest dimension, and renormalization group invariant at least to the leading order $\alpha_s$ in the quarkless world. It is noticed that the current $O_p(x)$ is anti-hermitian due to the involving of the imaginary time, while its analytic continuation to the Minkowskian space-time is hermitian. The QCD correlation function is defined as
\begin{equation}
\Pi(q^{2})=\int d^4x e^{ iq\cdot{x}  } \langle\Omega|T\{O_{p}(x)O^{\dag}_{p}(0)\}|\Omega\rangle .
\label{eq2:3}
\end{equation}
where $O^{\dag}_{p}$ is the hermitian conjugation of $O_{p}$. The advantages to use the hermitian conjugation in definition are: first, the spectral functions both in Euclidean space-time and in its analytic continuation into Minkowskian space-time are, in principle, positively definite; and second, the relationship between the correlation functions both in Euclidean and Minkowskian formulations becomes very simple as
\begin{eqnarray}
\Pi_E(Q^2=q^2)&\leftrightarrow \Pi_M(Q^2=-q^2),
\label{eq2:4}
\end{eqnarray}
because the overall minus sign arising from the analytic continuation due to
$(\epsilon_{\mu\nu\rho\sigma}\epsilon_{\mu\nu\rho\sigma})_E=-(\epsilon_{\mu\nu\rho\sigma}\epsilon^{\mu\nu\rho\sigma})_M$ is just canceled with another minus sign arising from the mentioned different hermiticity of the pseudoscalar glueball current. Therefore, the expressions for $\Pi_E(Q^2)$ and $\Pi_M(Q^2)$ are, in fact, the same function of $Q^2$.

In the framework of semiclassical expansion, the glue potential field $B(x)$ can be decomposed into a summation of the classical instanton $A$ and the corresponding quantum gluon field $a$ as
\begin{eqnarray}
B_{\mu}(x)=A_{\mu}(x)+a_{\mu}(x),
\label{eq2:5}
\end{eqnarray}
Consequently, the pure-glue Euclidean action can be expressed as
\begin{eqnarray}
  S[B]&=&S_0-\int\mathrm{d}^4x\,\left\{
  L[A+a]+\frac{1}{2\xi}a_{\mu}^a D_{\mu}^{ab}D_{\nu}^{bc} a_{\nu}^c \right\}
       \nonumber  \\
  &=&S_0-\frac{1}{2}\int\mathrm{d}^4x\,\left\{a_{\mu}^a\left[D_{\lambda}^{ab}D_{\lambda}^{bc}
  \delta_{\mu \nu}+2gf^{abc}F^b_{\mu \nu}\right.\right.     \nonumber  \\
  & &-\left.\left(1-\frac{1}{\xi}\right)D_{\mu}^{ab}D_{\nu}^{bc}\right]a_{\nu}^c
  -2gf_{abc}a_{\mu b}a_{\nu c}D_{\mu,ad}a_{\nu d}
                        \nonumber  \\
   & & -\left.\frac{1}{2}g^2f_{abc}a_{\mu b}a_{\nu c}f_{ade}a_{\mu d}a_{\nu e}\right\}
    \label{eq2:6}
\end{eqnarray}
where $S_0=8\pi^2/g^2$ is the one-instanton contribution to the
action, $F_{\mu \nu a}$ is the instanton field strength tensor
\begin{equation}
 F_{\mu\nu,a}(A)=\partial_{\mu}A_{\nu,a}-\partial_{\nu}A_{\mu,a}+g_sf_{abc}A_{\mu,b}A_{\nu,c},
  \label{eq2:7}
\end{equation}
and $D_{\mu}^{ab}(A)$ the covariant derivative associated with the
classical instanton field $A_{\mu}^a$
\begin{equation}
  D_{\mu}^{ab}(A)=\partial_\mu\delta_{ab}+gf_{acb}A_\mu^c
  \label{eq2:8}
\end{equation}
In addition, the background field gauge
\begin{equation}
  D^{\mu}(A)a_{\mu}=0
  \label{eq2:9}
\end{equation}
is used with $\xi$ being the corresponding gauge parameter, and certainly, the corresponding Faddeev-Popov ghosts according to the standard rule should be added to restore the unitarity. We note here that the structure constants $f_{abc}$ should be understood as $\epsilon_{abc}$ when any one of the color-indices $a$,$b$ and $c$ is associated with an instanton field due to the property of the closure of any group.

According to the decomposition \eqref{eq2:5}, the correlation function $\Pi$ splits into three parts, namely the pure classical part, the pure quantum part and the interference part in the leading order
\begin{equation}
\Pi^{\textrm{QCD}}(Q^2)=\Pi^{(\textrm{cl})}(Q^{2})+\Pi^{(\textrm{qu})}(Q^{2})+\Pi^{(\textrm{int})}(Q^{2}).
\label{eq2:10}
\end{equation}
where the superscript indicates that it is calculated in the underlying dynamical theory, QCD. It is important to note that every part in rhs of \eqref{eq2:10} is gauge-invariant because the decomposition \eqref{eq2:5}, in principle, has no impact on the gauge-invariance of the correlation function. The pure instaton contribution $\Pi^{(\textrm{cl})}(Q^{2})$ and the perturbative contribution $\Pi^{(\textrm{qu})}(Q^{2})$ up to three-loop level in the chiral limit of $\textrm{QCD}$ are shown in Eqs. \eqref{eqA:1} and \eqref{eqA:2} respectively in Appendix \ref{A}. The contribution of the topological charge screening which is not included in rhs of  \eqref{eq2:10} is given in Appendix \ref{B}.

One of our main tasks in this work is to calculate the contribution $\Pi^{(\textrm{int})}(Q^{2})$ in \eqref{eq2:10}, which is  arising from the interference between the classical instantons and quantum gluons in the framework of the semi-classical expansion for $\textrm{QCD}$ with the instanton background. After imposing the background covariant Feynman gauge ($\xi =1$) for the quantum gluon fields, we are still free to choose a gauge for the background field $A$. In the following, the singular gauge is chosen to the non-perturbative instanton field configurations as
\begin{equation}
A_{\mu{a}}(x)=\frac{2}{g_{s}}\eta_{a\mu\nu}(x-z)_{\nu}\phi(x-z),
\label{eq2:11}
\end{equation}
with
\begin{equation}
\phi(x-z)=
\frac{\rho^2}{(x-z)^2[(x-z)^2+\rho^2]},
\label{eq2:12}
\end{equation}
and the corresponding field strength tensor is
 \begin{eqnarray}
F_{\mu\nu,a}(x)&&=-\frac{8}{g_{s}}\left[\frac{(x-z)_{\mu}(x-z)_{\rho}}{(x-z)^2}-\frac{1}{4}\delta_{\mu\rho}\right]\notag\\
&&\times\eta_{a\nu\rho}\frac{\rho^2}{((x-z)^2+\rho^2)^2}-(\mu\leftrightarrow\nu),
\label{eq2:13}
\end{eqnarray}
with $z$ and $\rho$ denote respectively the center and size of the instanton, called collective coordinates together with the color orientation, and $\eta_{a\mu\nu}$ is the 't Hooft symbol which should be replaced with the anti-'t Hooft one $\bar{\eta}_{a\mu\nu}$ for an anti-instanton field. For the sake of simplicity, in the practice the most used is the spike size distribution $n(\rho)=\bar{n}\delta(\rho-\bar{\rho})$, where $\bar{n}$ is the overall instanton density and $\bar{\rho}$ is the average instanton size. The fact that the strong coupling constant $g_s$ emerging in the denominator of the rhs of \eqref{eq2:11} reveals the nonperturbative nature of these classical configurations. In fact, instantons play quite important role in $\textrm{QCD}$ sum rule. Early $\textrm{QCD}$ sum rules neglecting instanton-induced continuum contributions didn't obtain reliable results in many cases, but they are then solved by including such instanton-induced effects \cite{PhysRevD.64.034015,PhysRevLett.75.1707}.

Before starting with the contraction between the quantum fields, we note that the time-development of the instanton vacuum produces the pre-exponential factor for the distribution of the instantons\cite{Phys.Rev.Lett.37.8,PhysRevD.14.3432,PhysRevD.18.2199.2}, and $\Pi^{(\textrm{int})}$ is understood as taking ensemble average over the collective coordinates besides taking the usual vacuum expectation value due to the separation \eqref{eq2:5}
\begin{eqnarray}
&&\Pi^{(\textrm{int})}(x)=\notag\\
&&\sum_{I,\bar{I}}\int{d\rho}n(\rho)\int{d^4z} \langle\Omega|T\{O_{p}(x)O^{\dag}_{p}(0)\}^{(\textrm{int})}|\Omega\rangle,
\label{eq2:14}
\end{eqnarray}
where the super index '(\textrm{int})' indicates the corresponding quantity containing only the interference part between the quantum and classical ones. Using the spike distribution, \eqref{eq2:14} becomes
\begin{equation}
\Pi^{(\textrm{int})}(x)=2\bar{n}
\int{d^4z} \langle\Omega|T\{O_{p}(x)O^{\dag}_{p}(0)\}^{(\textrm{int})}|\Omega\rangle,
\label{eq2:15}
\end{equation}
where the factor 2 comes from the mutually equal contributions of both instanton and anti-instanton. Next important step is to specify the form of the gluon propagator which in the background field Feynman gauge can be read from the part of $S[B]$ quadratic in $a$\cite{Nucl.Phys.B199.451,Nucl.Phys.B201.141}
\begin{eqnarray}
{\cal D}_{\mu \nu}^{ab}(x,y)&=&\langle\Omega|T\{a^a_{\mu}(x)a^b_{\nu}(y)\}|\Omega\rangle\notag\\
&=&\langle x|\left(\frac{1}{P^2\delta_{\mu \nu}-2F_{\mu \nu}}\right)^{ab}|y\rangle
\label{eq2:16}
\end{eqnarray}
with $P^{ab}_{\mu}=-iD^{ab}_{\mu}$. Keeping only terms proportional to $F$, one has\cite{Fortsch.Phys.32.585}
\begin{align}
\int d^4x e^{iq\cdot x}{\cal D}_{\mu \nu}^{ab}(x,0)=&e^{iq\cdot (y-z)}\left\{
\frac{1}{q^2}\delta_{\mu \nu}+g_s\frac{2}{q^4}F_{\mu \nu}(z)\notag\right.\\
-&\left.ig_s\frac{(y-z)_{\rho}F_{\rho \sigma}(z)q_{\sigma}}{q^4}\delta_{\mu \nu}(z)+\cdots \right\}
\label{eq2:17}
\end{align}
where the first term in rhs of the above equation is the pure-gluon propagator in the usual Feynman gauge, and the second and third ones are the leading contribution of the instanton field to the gluon propagator. For short distance region, we assume that the contribution from a single instanton is dominant over multi-instantons\cite{Phys.Rev.D64.114020}. At the leading loop level, the gluon propagator \eqref{eq2:17} becomes the pure-gluon one.

\begin{figure}[!htbp]
\begin{center}
\includegraphics[width=3cm]{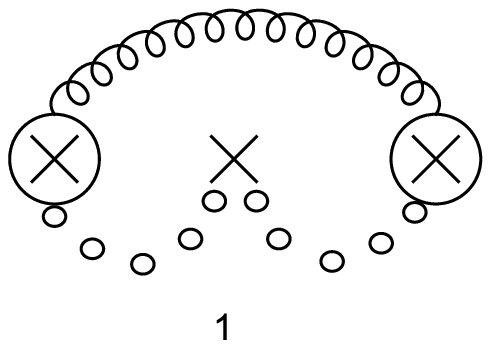}\qquad
\includegraphics[width=3cm]{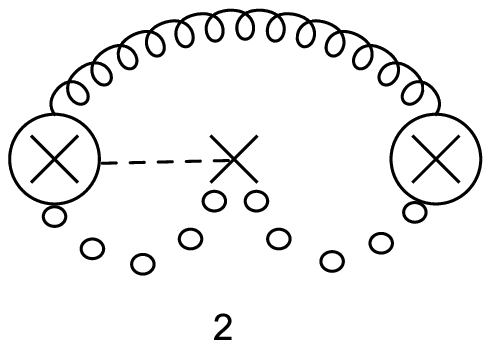}\qquad
\includegraphics[width=3cm]{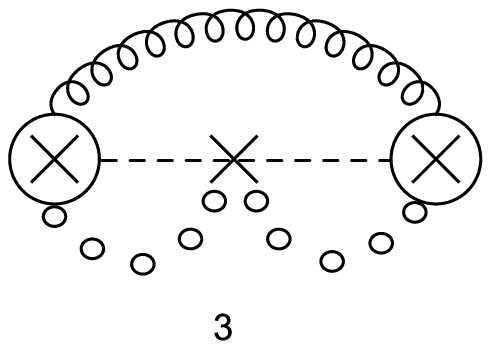}\qquad
\includegraphics[width=3cm]{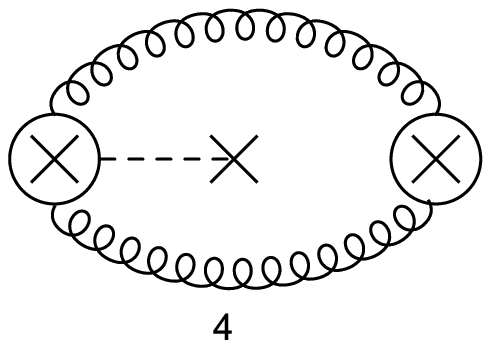}\qquad
\includegraphics[width=3cm]{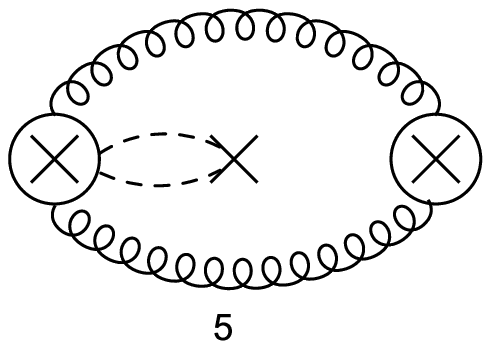}\qquad
\includegraphics[width=3cm]{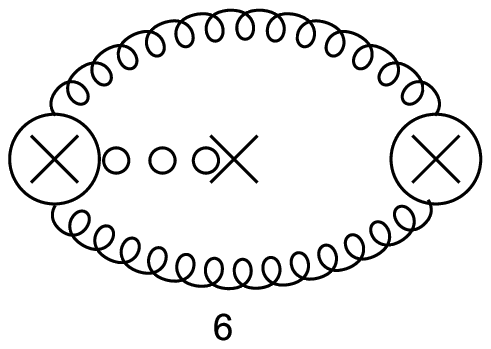}\qquad
\includegraphics[width=3cm]{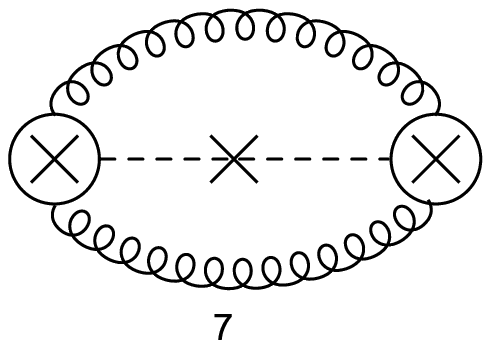}\qquad
\includegraphics[width=3cm]{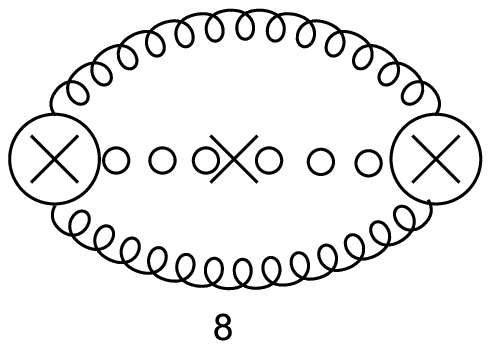}\qquad
\includegraphics[width=3cm]{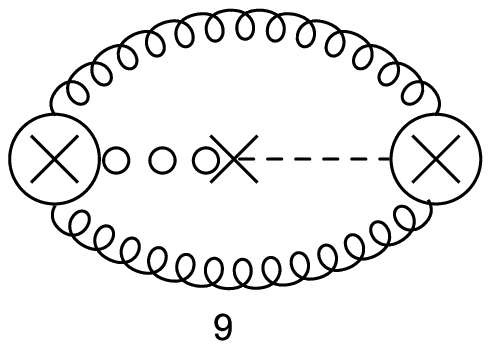}\qquad
\includegraphics[width=3cm]{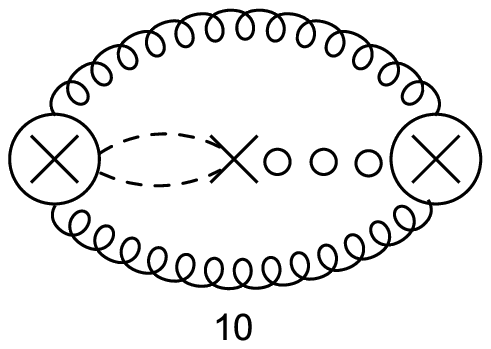}\qquad
\includegraphics[width=3cm]{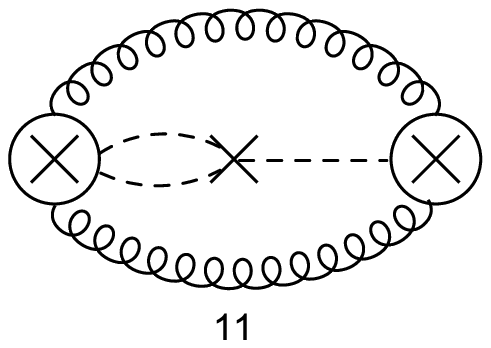}\qquad
\includegraphics[width=3cm]{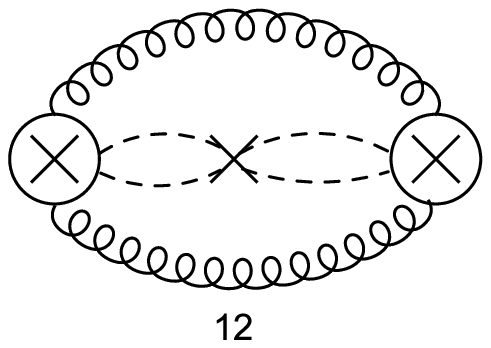}\qquad
\end{center} \caption{\small Feynman diagrams for the interference contribution $\Pi^{(\textrm{int})}(Q^{2})$ up to order $\alpha^2_s$, where spiral lines, dotted lines and the lines with circles denote gluons, instantons and the instanton
field strength tenser respectively, and cross stands for the position of instantons.}
\label{Fig:1}
\end{figure}

In calculation, we expand the current $O_p$ into terms which are the products of quantum gluon fields and their derivatives with coefficients being composed of the instanton fields
\begin{eqnarray}
O_P(x)&=&\frac{1}{2}\epsilon_{\mu\nu\rho\sigma}\alpha_s\sum^{10}_{i=1}O_i(x)
\label{eq2:18}
\end{eqnarray}
where the operators $O_i$ in terms of instanton and quantum gluon fields are listed in Appendix \ref{C}. Eq.\eqref{eq2:15} can be rewritten as
\begin{align}
\Pi^{(\textrm{int})}(q^2)=&-\frac{1}{2}\alpha^2_s\bar{n}\epsilon_{\mu\nu\rho\sigma}
\epsilon_{\mu'\nu'\rho'\sigma'}\notag\\
\times&\sum_{i,j}\int{d^4z} \int d^4x
e^{iq\cdot x} \langle\Omega|T\{O_i(x)O_j(0)\}|\Omega\rangle \nonumber  \\
=&\sum_{i=1}^{12}\Pi_i^{(\textrm{int})}(q^2)+\cdots
\label{eq2:19}
\end{align}
where the $\cdots$ denotes the contributions from the products of operators being proportional to $g_s^3$, and the expressions of $\Pi_i^{(\textrm{int})}(q^2)$ in terms of $O_i$ are shown in Appendix \ref{D}. The corresponding twelve kinds of Feynman diagrams as shown in the FIG. \ref{Fig:1}, where the contributions from the first three diagrams are of the order of $\alpha_s$, and the contributions of the remainders are superficially of the order of $\alpha^2_s$, and those from the diagrams (4),(5) and (6), in fact, are vanishing because of violating the conservation of the color-charge, namely
\begin{equation}
\Pi_i^{(\textrm{int})}(q^2)=0,\hspace{1cm}\textrm{for }i=4,5,6
\label{eq2:20}
\end{equation}

Now, we are in the position to evaluate the contributions of the remainder diagrams in FIG. \ref{Fig:1}. Using the standard technique to regularizing the ultraviolet divergence in the modified minimal subtraction scheme, the result for the interference part of the correlation function is
\begin{eqnarray}
\Pi^{\textrm{int}}(Q^2)&=&c_0\alpha_s\bar{n}\pi+\alpha_s^2\bar{n}\left\{c_1+c_2(Q\rho)^{-2}\right.\nonumber\\
&+&\left.\left[c_3(Q\rho)^2+c_4+
c_5(Q\rho)^{-2}\right]\ln\frac{Q^2}{\mu^{2}}\right\},
\label{eq2:21}
\end{eqnarray}
where we have ignored terms being proportional to the positive powers of $q^2$ which vanish after Borel transformation, and the dimensionless coefficients $c_i$ are numerically determined to be
\begin{align}
 &c_0=-118.23,\,\,  c_1=-3700.59\alpha_s,\,\,\, c_2=-2394.47\alpha_s,\,\nonumber\\
 & c_3=11561.90\alpha_s,\,\, c_4=1850.30\alpha_s,\,\, c_5=1197.24\alpha_s.
 \label{eq2:22}
\end{align}
through a tedious calculation. It should be noted that there is no infrared divergence as expected by the instanton size being fixed in the liquid instanton vacuum model. Comparing Eq.\eqref{eq2:21} with our previous result \cite{Eur.Phys.J.Plus128.115,J.Phys.G.41.035004}, they differ not only in some coefficients but also in the logarithms structures due to the fact that the newly improved calculation is free of infrared divergence while our old one were not, and it would need a corresponding cutoff to regularize the integral in the infrared limit.

Putting everything above together, our final correlation function for the pseudoscalar glueball current is of the form
\begin{eqnarray}
\Pi^{\textrm{QCD}}(Q^2)&=&-2^5\pi^2\bar{n}y^4K^2_2(y)\notag\\
&+&\alpha_s\bar{n}\left[c_0\pi+c_1+c_2y^{-2}\right.\notag\\
&+&\left.\left(c_3y^2+c_4+c_5y^{-2}\right)\ln\frac{Q^2}{\mu^{2}}\right]\nonumber\\
&+&\left(\frac{\alpha_s}{\pi}\right)^2Q^4\ln\frac{Q^2}{\mu^2}\left[a_0+ a_1\ln\frac{Q^2}{\mu^2}\right.\notag\\
&+&\left.a_2\ln^2\frac{Q^2}{\mu^2}\right].
 \label{eq2:23}
\end{eqnarray}
with $y=Q\bar{\rho}$.

\begin{figure}[!hbt]
\centering
\includegraphics[width=8.6cm]{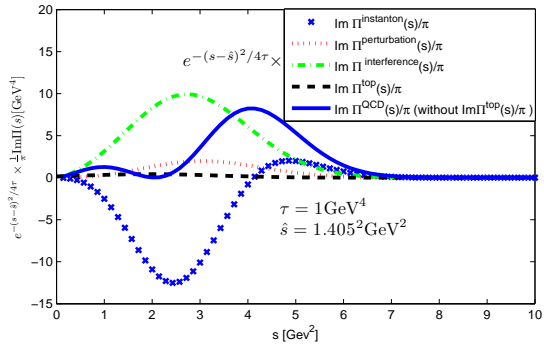}
\caption{\small The contributions to the imaginary part of the correlation function from the pure instanton (cross line), interference (dashed-dotted line), pure perturbative (dotted line) and topological charge screening (dashed line) and the total contribution without the topological charge screening one (solid line) versus $s$.}
\label{Fig:2}
\end{figure}

Before going on, let us compare the roles of the various parts of the contributions in correlation function.
The imaginary part of the correlation function \eqref{eq2:23} can be worked out to be
\begin{eqnarray}
\frac{1}{\pi}\textrm{Im}\Pi^{\textrm{QCD}}(s)&=&
16\pi^3s^2\bar{n}\bar{\rho}^4J_2(\bar{\rho}\sqrt{s})Y_2(\bar{\rho}\sqrt{s})\nonumber\\
&&+\alpha_s^2\bar{n}\left[c_3\bar{\rho}^2s-c_4+c_5(\bar{\rho}^2s)^{-1}\right]\nonumber\\
&&-s^2\left( \frac{\alpha_s}{\pi}\right)^2\left[a_0+2a_1\ln \frac{s}{\mu^2}\right.\notag\\
&&\left.+\left(3\ln^2 \frac{s}{\mu^2}-\pi^2\right)a_2\right]
\label{eq2:24}.
\end{eqnarray}
where the pure classical contribution (the first term on rhs of \eqref{eq2:24}) is most dominant, and the contribution of the interference terms (the second term on rhs) is of the second place, the pure perturbative contribution simply plays the role of the third place, as shown in FIG. \ref{Fig:2} where the imaginary part of the correlation function is multiplied with a weight function $\exp{(-(s-\hat{s})/4\tau)}$ as required by the Gaussian sum rules, and in accordance with the spirit of the semiclassical expansion. All these three contributions are positively definite as expected. We note that the contribution from the topological charge screening, from \eqref{eqB:2}, is displayed in FIG. \ref{Fig:2} as well, and its role is almost insignificant. Moreover, it is easy to see from FIG. \ref{Fig:2} that the imaginary part of the correlation function is already positive from $s=0.5\mathrm{GeV}^2$ to $s=10\mathrm{GeV}^2$ without including the contribution of the topological charge screening, and the so-called positivity problem is no longer there. Therefore, the interference contribution to the correlation function is significant important not only in its magnitude but also in restoring the positivity to the spectral function.

We note here that the so-called condensate contribution to the correlation function of the pseudoscalar glueball current is proven to be very small in comparing with the one of \eqref{eq2:23}, as shown in Appendix \ref{E}, where the comparison between the real and imaginary parts of the correlation function and condensate contribution to it are made, and shown in FIG. \ref{Fig:8} and \ref{Fig:9}. For the reasons given above, the contributions from the topological charge screening effect and the usual condensates are omitted in our sum rule analysis.

\section{Spectral function}
\label{sectionIII}
In the isosinglet channel there are five gauge-invariant composite operators with the quantum numbers of $0^{-+}$ which  are bilinear in the fundamental quark, antiquark and gluon fields, namely the pseudoscalar quark densities, the divergences of the axial quark currents and the gluon anomaly:
\begin{equation}
\hat{J}^{8,0}_5=i\bar{q}\gamma_5(\lambda^{8,0}/2)q,
\label{eq4:25}
\end{equation}
\begin{equation}
\partial_\mu{\hat{J}^{8,0}_{\mu5}}=\partial_{\mu}[\bar{q}\gamma_\mu\gamma_5(\lambda^{8,0}/2)q],
\label{eq4:26}
\end{equation}
\begin{equation}
\hat{O}_p=\alpha_sG^a_{\mu\nu}\widetilde{G}^a_{\mu\nu},
\label{eq4:27}
\end{equation}
where $\lambda^{8}$ is the flavor Gell-Mann matrix, and $\lambda^{0}=\sqrt{2/3}I$ with $I$ being the $3\times3$ flavor unit matrix. Only three of these operators are independent due to the two renormalization-invariant axial Ward identities
\begin{eqnarray}
\partial_\mu{\hat{J}^{8}_{\mu5}}&=&\frac{1}{3}(m_u+m_d+4m_s)\hat{J}^8_5\nonumber\\
&&+\frac{1}{3}(4m_u+4m_d+2m_s)\hat{J}^0_5,
\label{eq4:28}
\end{eqnarray}
\begin{eqnarray}
\partial_\mu{\hat{J}^{0}_{\mu5}}&=&\frac{\sqrt{2}}{3}(m_u+m_d-2m_s)\hat{J}^8_5\nonumber\\
&&+\frac{2}{3}(m_u+m_d-2m_s)\hat{J}^0_5+\frac{\sqrt{3}}{4\sqrt{2}\pi}\hat{O}_p.
\label{eq4:29}
\end{eqnarray}
Further, under renormalization and the flavor space rotation, we have
\begin{equation}
(m\hat{J}^{8,0}_{5})_{(r)}=m\hat{J}^{8,0}_5,
\label{eq4:30}
\end{equation}
\begin{equation}
(\partial_\mu{\hat{J}^{8}_{\mu5}})_{(r)}=\partial_\mu{\hat{J}^{8}_{\mu5}},(\partial_\mu{\hat{J}^{0}_{\mu5}})_{(r)}=Z\partial_\mu{\hat{J}^{0}_{\mu5}},
\label{eq4:31}
\end{equation}
\begin{equation}
(\hat{O}_p)_{(r)}=\hat{O}_p+\frac{4\sqrt{2}\pi}{\sqrt{3}}\partial_\mu{\hat{J}^0_{\mu5}},
\label{eq4:32}
\end{equation}
where the quantities with subscript $(r)$ are the renormalized ones. As a consequence, the gluon anomaly operator $\hat{O}_p$ even through renormalization-invariant in the pure gluon world, is a linear combination of three operators $\hat{J}^8_5$, $\hat{J}^0_5$ and $\hat{O}_p$ after renormalization. Therefore, one assumes that there may be some isosinglet quark-antiquark pseudoscalar states mixed with the pseudoscalar glueball ground state G.

Now we construct the spectral function for the correlation function of the pseudoscalar glueball current.
The usual lowest one resonance plus a continuum model is used to saturate the phenomenological spectral function:
\begin{equation}
\frac{1}{\pi}\textrm{Im}\Pi^{\textrm{PHEN}}(s)
=\rho^{\textrm{HAD}}(s)-\theta(s-s_{0})\frac{1}{\pi}\textrm{Im}\Pi^{\textrm{QCD}}(s),
 \label{eq4:33}
\end{equation}
where $s_0$ is the QCD-hadron duality threshold, $\theta(s-s_{0})$ the step function and  $\rho^{\textrm{HAD}}(s)$ the spectral function for the lowest pseudoscalar glueball state.
In the usual zero-width approximation, the spectral function for a single resonance is assumed to be
\begin{equation}
\rho^{\textrm{HAD}}(s)=F^2\delta(s-m^2),
 \label{eq4:34}
\end{equation}
where $m$ is the mass of the lowest glueball, and $F$  is the coupling constant of the current to the glueball defined as
\begin{equation}
\langle0|O_{p}(0)|G\rangle=F.
\label{eq4:35}
\end{equation}
The threshold behavior for $\rho^{\textrm{HAD}}(s)$ is known to be
\begin{equation}
\rho^{\textrm{HAD}}(s)\rightarrow\lambda_0s , ~\textrm{for} ~s\rightarrow0
 \label{eq4:36}
\end{equation}
from the low-energy theorem in the world of no light quark flavors \cite{NuclPhysB.191.301} or the one in the world with three light flavors and $m_{u,d}\ll m_s$\cite{Phys.Rev.D46.5607}. In fact, the threshold behavior \eqref{eq4:36} is only proven to be valid near by the chiral limit; it may not be extrapolated far away. Therefore, instead of considering the coupling $F$ as a constant \cite{PhysRevD.71.054008}, we choose a
model for $F$ as
\begin{equation}
 F=
 \left\{
 \begin{array}{ll}
  \lambda_0s,        &\mbox{for } s<m^2_{\pi},\\
  fm^2,&\mbox{for } s\geq m^2_{\pi},
 \end{array}\right.
 \label{eq4:37}
\end{equation}
where the $\lambda_0$ and $f$ are some constants determined late in numerical simulation.

To go beyond the zero-width approximation, in facing the near-actual situation, the Breit-Wigner form for a single resonance is assumed for $\rho^{\textrm{HAD}}(s)$
\begin{equation}
\rho^{\textrm{HAD}}(s)=\frac{ F^2m\Gamma}{(s-m^2+\Gamma^2/4)^2+m^2\Gamma^2},
 \label{eq4:38}
\end{equation}
where $\Gamma$ is the width of the lowest glubeball.

Further, the one isolated lowest resonance assumption is questioned from the admixture with quarkonium states, and it is known from the experimental data that there are five $0^{-+}$ pseudoscalar resonances till and around the mass scale of 1.405 $\textrm{GeV}$ (namely $\eta(548)$, $\eta(958)$, $\eta(1295)$, $\eta(1405)$ and $\eta(1475)$). The form of the spectral function for five resonances is taken to be
\begin{equation}
\rho^{\textrm{HAD}}(s)=\sum^5_{i=1}\frac{ F^2_im_i\Gamma_i}{(s-m^2_i+\Gamma^2_i/4)^2+m^2_i\Gamma^2_i},
 \label{eq4:39}
\end{equation}
where $m_i$ and $\Gamma_i$ being the mass and width of the i-th resonance, respectively. For the sake of simplicity, all coupling constants $F_i$ for $s<m^2_{\pi}$ are fixed with the same $\lambda_0$ as shown in \eqref{eq4:37}.

It is noticed that there are other pseudoscalar resonances $\eta(1760)$ and $\eta(2225)$ which are omitted from the summary table of PDG. These two resonances are excluded in our consideration. The reasons may be listed in order: First, although $\eta(1760)$ and $\eta(2225)$ may certainly be coupled to the pseudoscalar glueball current via the gluon anomaly, such coupling, however, contains a factor of the running coupling as commonly seen in QCD, and becomes weaker with $Q^2$ increases, as demonstrated in an effective QCD low energy theory \cite{Phys.Rev.Lett80.434}.
Second, the mixing between the considered pseudoscalar glueball and $\eta(1760)$ and $\eta(2225)$ is believed to be very small because the locations of $\eta(1760)$ and $\eta(2225)$ are far away from the scale of the lowest pseudoscalar glueball. Third, the continuum threshold $s_0$, determined in our sum rule approach, is only in an effective sense due to the accuracy level of the present calculation.

\section{Finite width Gaussian sum rules}
\label{sectionIV}
In this section, we construct the appropriate sum rules of $0^{-+}$ pseudoscalar glueball current which has the form due to dispersion relation
\begin{equation}
\Pi^{\textrm{QCD}}(Q^2)=\int^{\infty}_0ds\frac{1}{s+Q^2}\frac{1}{\pi}\textrm{Im}\Pi(s).
 \label{eq4:40}
\end{equation}
where $\textrm{Im}\Pi(s)$ could be simulated by the phenomenological one $\textrm{Im}\Pi^{\textrm{PHEN}}(s)$  whithin an assumed model of the spectral function in a spirit of sum rule approach. Using the Borel transformation \cite{NuclearPhysicsB.250.61}
\begin{equation}
\hat{\mathcal{B}}\equiv{\lim_{\left.\begin{subarray}{c}N\rightarrow\infty\\Q^4\rightarrow
\infty\end{subarray}\right|}}_{Q^4/N\equiv4\tau}
\frac{(-1)^N}{(N-1)!}\,(Q^4)^N\left(\frac{\textrm{d}}{\textrm{d}Q^4}\right)^N
\label{eq4:41}
\end{equation}
to both sides of \eqref{eq4:40}, a family of Gaussian sum rules can be formed to be \cite{NuclearPhysicsB.250.61}
\begin{eqnarray}
 \mathcal{G}^{\textrm{HAD}}_{k}(s_{0},\hat{s},\tau)&=&\mathcal{G}^{\textrm{QCD}}_{k}(s_{0},\hat{s},\tau)\nonumber\\
 &&+\frac{1}{\sqrt{4\pi\tau}}\exp[-\frac{\hat{s}^2}{4\tau}]\Pi(0)\delta_{k,-1},
\label{eq4:42}
\end{eqnarray}
where $s_0$ is the continuum threshold which hadronic physics is (locally) dual to $\textrm{QCD}$ above it.
\begin{eqnarray}
  \Pi(0)=(8\pi)^2\frac{m_u m_d}{m_u+m_d}\langle \bar{q}q \rangle
  \label{eq4:43}
\end{eqnarray}
comes from the low-energy theorem for QCD with three light flavors\cite{Phys.Rev.D46.5607}, and
\begin{equation}
\mathcal{G}^{\textrm{HAD}}_{k}(s_{0},\hat{s},\tau)=\frac{1}{\sqrt{4\pi\tau}}
\int^{s_{0}}_{0}dss^{k}\exp[-\frac{(s-\hat{s})^2}{4\tau}]\frac{\rho^{\textrm{HAD}}(s)}{\pi},
\label{eq4:44}
\end{equation}
for the phenomenological contributions to the sum rules, and
\begin{equation}
 \mathcal{G}^{\textrm{QCD}}_{k}(s_{0},\hat{s},\tau)
 =\mathcal{G}^{\textrm{QCD}}_{k}(\hat{s},\tau)-\mathcal{G}^{\textrm{CONT}}_{k}(s_{0},\hat{s},\tau),
\label{eq4:45}
\end{equation}
for the theoretical contributions, where $\mathcal{G}^{\textrm{CONT}}_{k}(s_{0},\hat{s},\tau)$ is the contribution of continuum being defined as
\begin{equation}
\mathcal{G}^{\textrm{CONT}}_{k}(s_{0},\hat{s},\tau)
=\frac{1}{\sqrt{4\pi\tau}}\int^{\infty}_{s_{0}}dss^{k}
\exp[-\frac{(s-\hat{s})^2}{4\tau}]\frac{\textrm{Im}^{\textrm{QCD}}(s)}{\pi},
\label{eq4:46}
\end{equation}
and $\mathcal{G}^{\textrm{QCD}}_{k}(\hat{s},\tau)$ is defined as
\begin{eqnarray}
\mathcal{G}^{\textrm{QCD}}_{k}(\hat{s},\tau)&=&
\frac{2\tau}{\sqrt{4\pi\tau}}\mathcal{\hat{B}}\left[\frac{(\hat{s}
+iQ^2)^{k}\Pi^{\textrm{QCD}}(\hat{s}+iQ^2)}{iQ^{2}}\right.\notag\\
&-&\left.\frac{(\hat{s}-iQ^2)^{k}\Pi^{\textrm{QCD}}(\hat{s}-iQ^2)}{iQ^{2}}\right].
\label{eq4:47}
\end{eqnarray}

Substituting the correlation function \eqref{eq2:23} of $0^{-+}$ pseudoscalar glueball into \eqref{eq4:47}, one can derive the Gaussian sum rules of $k=-1,0$ and $+1$

\begin{eqnarray}
\mathcal{G}^{\textrm{QCD}}_{-1}(\hat{s},\tau)
&=&\hat{\mathcal{I}}\cdot16\pi^3\bar{n}\bar{\rho}^4J_2\left(\bar{\rho}\sqrt{s}\right)Y_2\left(\bar{\rho}\sqrt{s}\right)s\nonumber\\
&+&\bar{n}\pi\alpha_sc_0\frac{1}{\sqrt{4\pi\tau}}\exp\left[-\frac{\hat{s}^2}{4\tau}\right]\notag\\
&+&\bar{n}\alpha_s^2\frac{1}{\sqrt{4\pi\tau}}\left\{-(c_1+c_3\gamma)\exp\left[-\frac{\hat{s}^2}{4\tau}\right]\right.\notag\\
&+&c_2\bar{\rho}^2\sqrt{2\tau}e^{-\hat{s}^2/8\tau}D_{-1}(-\hat{s}/\sqrt{2\tau})\notag\\
&-&\left.(c_4/\bar{\rho}^2-c_5(\gamma-1)/\bar{\rho}^2)\frac{\hat{s}}{2\tau}\exp\left[-\frac{\hat{s}^2}{4\tau}\right]\right\}\notag\\
&+&\frac1{\sqrt{4\pi\tau}}D_{-2}(-\hat{s}/\sqrt{2\tau})[a_0-(2\gamma-2)a_1\notag\\
&+&0.5(6\gamma^2-12\gamma-\pi^2)a_2]2\tau e^{-\hat{s}^2/8\tau},
\label{eq4:48}
\end{eqnarray}

\begin{eqnarray}
\mathcal{G}^{\textrm{QCD}}_{0}(\hat{s},\tau)
&=&\hat{\mathcal{I}}\cdot16\pi^3\bar{n}\bar{\rho}^4J_2\left(\bar{\rho}\sqrt{s}\right)Y_2\left(\bar{\rho}\sqrt{s}\right)s^2\nonumber\\
&+&\bar{n}\alpha_s^2\frac1{\sqrt{4\pi\tau}}\left\{c_2\bar{\rho}^22\tau e^{-\hat{s}^2/8\tau}D_{-2}(-\hat{s}/\sqrt{2\tau})\right.\notag\\
&-&c_3\sqrt{2\tau}e^{-\hat{s}^2/8\tau}D_{-1}(-\hat{s}/\sqrt{2\tau})\notag\\
&+&\left.(c_4/\bar{\rho}^2-c_5\gamma/\bar{\rho}^2)\exp\left[-\frac{\hat{s}^2}{4\tau}\right]\right\}\notag\\
&+&\frac1{\sqrt{4\pi\tau}}D_{-3}(-\hat{s}/\sqrt{2\tau})[2a_0+(6-4\gamma)a_1\notag\\
&+&(6\gamma^2-18\gamma-\pi^2+6)a_2](2\tau)^{3/2} e^{-\hat{s}^2/8\tau},
\label{eq4:49}
\end{eqnarray}

\begin{eqnarray}
\mathcal{G}^{\textrm{QCD}}_{1}(\hat{s},\tau)
&=&\hat{\mathcal{I}}\cdot16\pi^3\bar{n}\bar{\rho}^4J_2\left(\bar{\rho}\sqrt{s}\right)
Y_2\left(\bar{\rho}\sqrt{s}\right)s^3\nonumber\\
&+&\frac{\bar{n}\alpha_s^2}{\sqrt{4\pi\tau}}\left\{2c_2\bar{\rho}^2(2\tau)^{3/2} e^{-\hat{s}^2/8\tau}D_{-3}(-\hat{s}/\sqrt{2\tau})\right.\notag\\
&-&c_3 2\tau e^{-\hat{s}^2/8\tau}D_{-2}(-\hat{s}/\sqrt{2\tau})\notag\\
&+&\left.c_5\bar{\rho}^{-2}\sqrt{2\tau}e^{-\hat{s}^2/8\tau}D_{-1}(-\hat{s}/\sqrt{2\tau})\right\}\notag\\
&+&\frac1{\sqrt{4\pi\tau}}D_{-4}(-\hat{s}/\sqrt{2\tau})[6a_0+(22-12\gamma)a_1\notag\\
&+&(18\gamma^2-66\gamma-3\pi^2+36)a_2]4\tau^2e^{-\hat{s}^2/8\tau},
\label{eq4:50}
\end{eqnarray}
where
\begin{align}
&&\hat{\mathcal{I}}=\frac{1}{\sqrt{4\pi\tau}}\int^{\infty}_{0}\textrm{d}s\exp\left[-\frac{(s-\hat{s})^2}{4\tau}\right],
\label{eq4:51}
\end{align}
and parabolic cylinder function $D_{-d-1}(\hat{s}\sqrt{\tau})$ defined as
\begin{align}
&&D_{-d-1}(z)=\frac{\sqrt{2}(-1)^d e^{-z^2/4}}{d!}\frac{\textrm{d}^d\left(e^{z^2/2}\int^{\infty}_{z/\sqrt{2}}\textrm{d}ye^{-y^2}\right)}{(\textrm{d}z)^d},\notag\\
&&d\geq0.
\label{eq4:52}
\end{align}

\section{Numerical simulation}
\label{sectionV}
The expressions for the three-loop running coupling constant $\alpha_{s}(Q^2)$ with three massless flavors $(N_{f}=3)$ at renormalization scale $\mu$~\cite{Prog.Part.Nucl.Phys.58.387}
\begin{equation}
\frac{\alpha_s(\mu^2)}{\pi}=\frac{\alpha^{(2)}_s(\mu^2)}{\pi}+\frac{1}{(\beta_0L)^3}\left[L_{1}(\frac{\beta_1}{\beta_0})^2+\frac{\beta_2}{\beta_0}\right]
\label{eq5:53}
\end{equation}
are used, where ${\alpha^{(2)}_s(\mu^2)}/{\pi}$ is the two-loop running coupling constant with $(N_{f}=0)$
\begin{equation}
\frac{\alpha^{(2)}_s(\mu^2)}{\pi}=\frac{1}{\beta_0L}-\frac{\beta_1}{\beta_0}\frac{\ln L}{(\beta_0 L)^2}
\label{eq5:54}
\end{equation}
and
\begin{eqnarray}
&&L=\ln\left(\frac{\mu^2}{\Lambda^2}\right), L_1=\ln^2{L}-\ln{L}-1,\quad \nonumber\\
&&\beta_0=\frac1{4}\left[11-\frac2{3}N_f\right],\quad \nonumber\\
&&\beta_1=\frac1{4^2}\left[102-\frac{38}{3}N_f\right],\quad\nonumber\\
&&\beta_2=\frac1{4^3}\left[\frac{2857}{2}-\frac{5033}{18}N_f+\frac{325}{54}N_f\right],
\label{eq5:55}
\end{eqnarray}
with the color number $N_c=3$ and the QCD renormalization invariant scale $\Lambda=120\textrm{MeV}$.
We take $\mu^2=\sqrt{\tau}$ after calculating Borel transforms based on the renormalization group improvement for Gaussian sum rules \cite{Physics.Letters.B.103.57}. The subtraction constant $\Pi(0)$ has been fixed as \cite{PhysRevD.71.054008}
\begin{equation}
\Pi(0)\simeq-0.022\textrm{GeV}^4,
\label{eq5:56}
\end{equation}
and the values of the average instanton size and the overall instanton density are adopted from the instanton liquid model\cite{NuclPhysB.203.116}
\begin{eqnarray}
&&\overline{n}=1\textrm{fm}^{-4}=0.0016 \textrm{GeV}^{4},\nonumber\\
&&\overline{\rho}=\frac{1}{3}\textrm{fm}\simeq1.667 \textrm{GeV}^{-1}.
\label{eq5:57}
\end{eqnarray}

\begin{figure}[hbt]
\centering
\includegraphics[width=8.6cm]{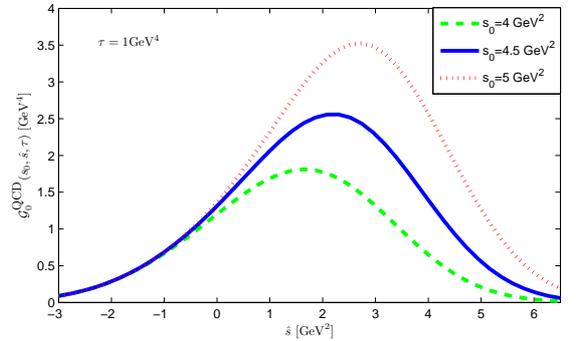}
\caption{\small The curves of $\mathcal{G}^{\textrm{QCD}}_{0}(s_0,\hat{s},\tau)$ versus $\hat{s}$ with different $s_0$ at $\tau=1~\textrm{GeV}^4$.}
\label{Fig:3}
\end{figure}

The resonance parameters in Eq.\eqref{eq4:39} could be estimated by matching both sides of sum rules Eq.\eqref{eq4:42} optimally in the fiducial domain. In doing so, the parameter $\hat{s}$ and  the threshold  $s_0$ should be determined priority. Firstly, it is obvious that $s_0$ must be greater than the mass
square of the highest lying isolated resonance considered, namely
\begin{align}
s_0\geq m_{\textrm{max}}^2
\label{eq5:58}
\end{align}
in our multi-resonance assumption (or just the resonance mass itself in the case of a single resonance assumption), and should guarantee that there is a sum rule window for Gaussian sum rules. Secondly, the peak positions of $\mathcal{G}^{\textrm{QCD}}_{k}(s_0,\hat{s},\tau)$ versus $\hat{s}$ curves should not change too much with moderate variation of $s_0$. Here we do not mention about the values of $\tau$ in this condition, because the peak positions of these curves is not affected by appropriate values of $\tau$. It is found that the behavior of these curves can satisfy above requirements as shown in FIG. \ref{Fig:3} if $s_0$ lies in the interval of $(4\textrm{GeV}^2,5\textrm{GeV}^2)$. It is also remarkable that the peak positions  has already indicated the approximate mass of the hadron considered. Thus, we would expect that the mass of the pseudoscalar glueball should be nearby the value $\sqrt{\hat{s}}\simeq 1.449 ~\textrm{GeV}$. In another way, if one uses the curves of $\mathcal{G}^{\textrm{QCD}}_{k}(s_0,\hat{s},\tau)$ versus $\tau$ with fixed $\hat{s}$ and $s_0$ to obtain the physical parameters through \eqref{eq4:43}, then $\hat{s}$ should be set approximately to be $\hat{s}_{\textrm{peak}}$ of the curves of $\mathcal{G}^{\textrm{QCD}}_{k}(s_0,\hat{s},\tau)$ versus $\tau$, so as to highlight the underlying hadron state in consideration and suppress the contributions from other states.
Besides, it needs a sum rule window which the hadron physical properties should be stable in this region. For the upper limit $\tau_{\textrm{max}}$ of the sum rule window, the resonance contribution should be great than the continuum one
\begin{equation}
\mathcal{G}^{\textrm{QCD}}_{k}(s_0,\hat{s},\tau_{\textrm{max}})\geq\mathcal{G}^{\textrm{CONT}}_{k}(s_0,\hat{s},\tau_{\textrm{max}})
\label{eq5:59}
\end{equation}
according to the standard requirement due to the fact that in the energy region above $\tau_{\textrm{max}}$ the perturbative contribution is dominant.
At $\tau_{\textrm{min}}$, which lies in the low-energy region, we require that the single instanton contribution should be relatively large so that
\begin{equation}
\frac{\mathcal{G}^{\textrm{inst}}_{k}(s_0,\hat{s},\tau_{\textrm{min}})}{\mathcal{G}^{\textrm{QCD}}_{k}(s_0,\hat{s},\tau_{\textrm{min}})}\geq50\%
\label{eq5:60}
\end{equation}
In the same time, to require that the multi-instanton corrections remain negligible, we simply adopt a rough estimate
\begin{equation}
\tau_{\textrm{min}}\geq 2\bar{\rho}^{-4}\simeq\left(\frac{2}{0.6\textrm{GeV}}\right)^4.
\label{eq5:61}
\end{equation}
 According to the above requirements, we find that in the domain
\begin{equation}
\tau\in(0.5,4.5)\textrm{GeV}^4,
\label{eq5:62}
\end{equation}
our sum rules work very well. Finally, in order to measure the compatibility between both sides of the sum rules \eqref{eq4:42} in our numerical simulation, we divide the sum rule window $[\tau_{\mathrm{min}},\tau_{\mathrm{max}}]$ into $N=100$ segments of equal width, $[\tau_i,\tau_{i+1}]$, with $\tau_0=\tau_{\mathrm{min}}$ and $\tau_N=\tau_{\mathrm{max}}$, and introduce a variation $\delta$ (called the matching measure) which is defined as
\begin{equation}
\delta=\frac{1}{N}\sum_{i=1}^N\frac{[L(\tau_i)-R(\tau_i)]^2}{|L(\tau_i)R(\tau_i)|},
\label{eq5:63}
\end{equation}
where $L(\tau_{i})$ and $R(\tau_{i})$ are lhs and rhs of \eqref{eq4:42} evaluated at $\tau_i$.

\begin{figure}[hbt]
\centering
\includegraphics[width=8.0cm]{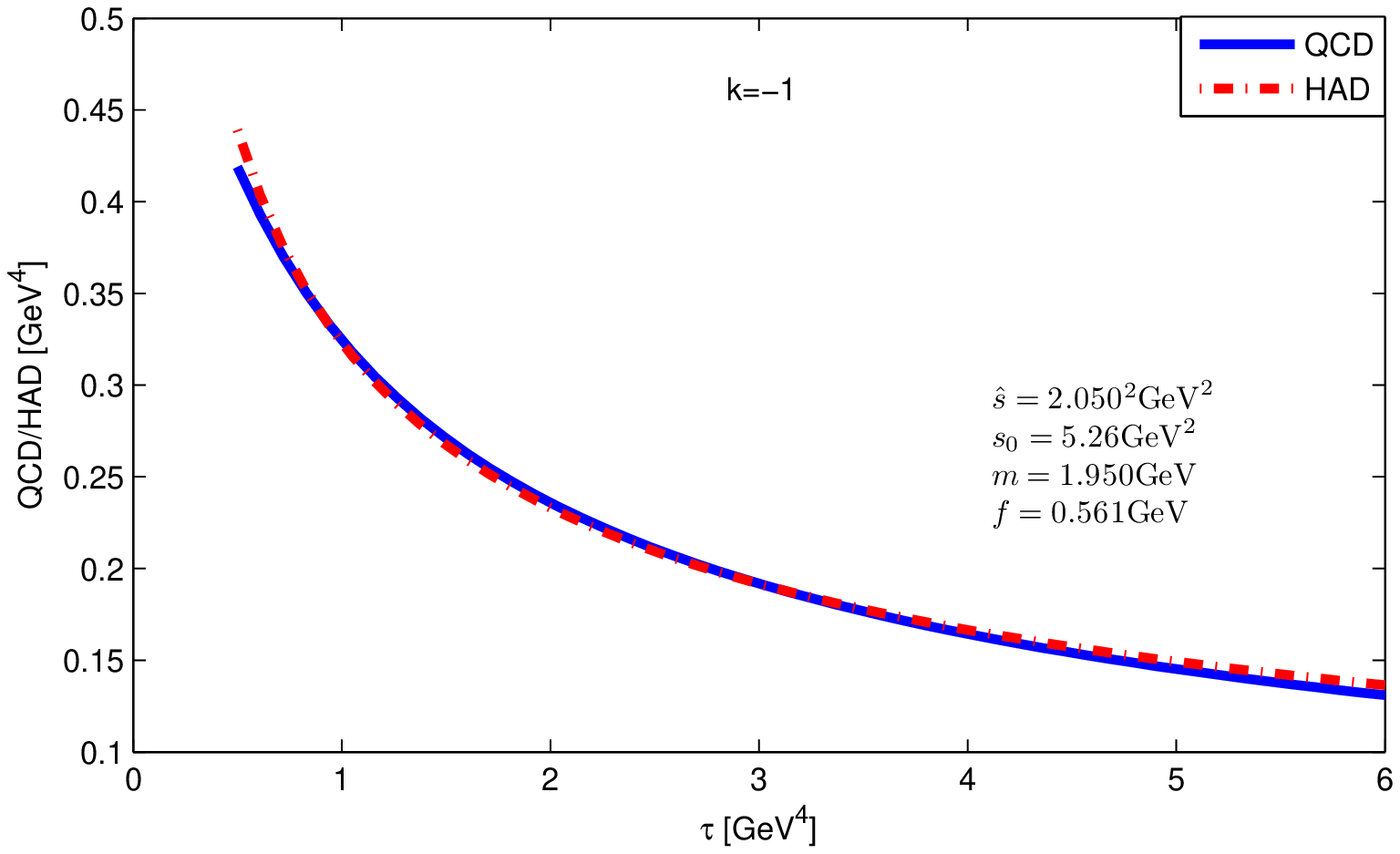}
\includegraphics[width=8.0cm]{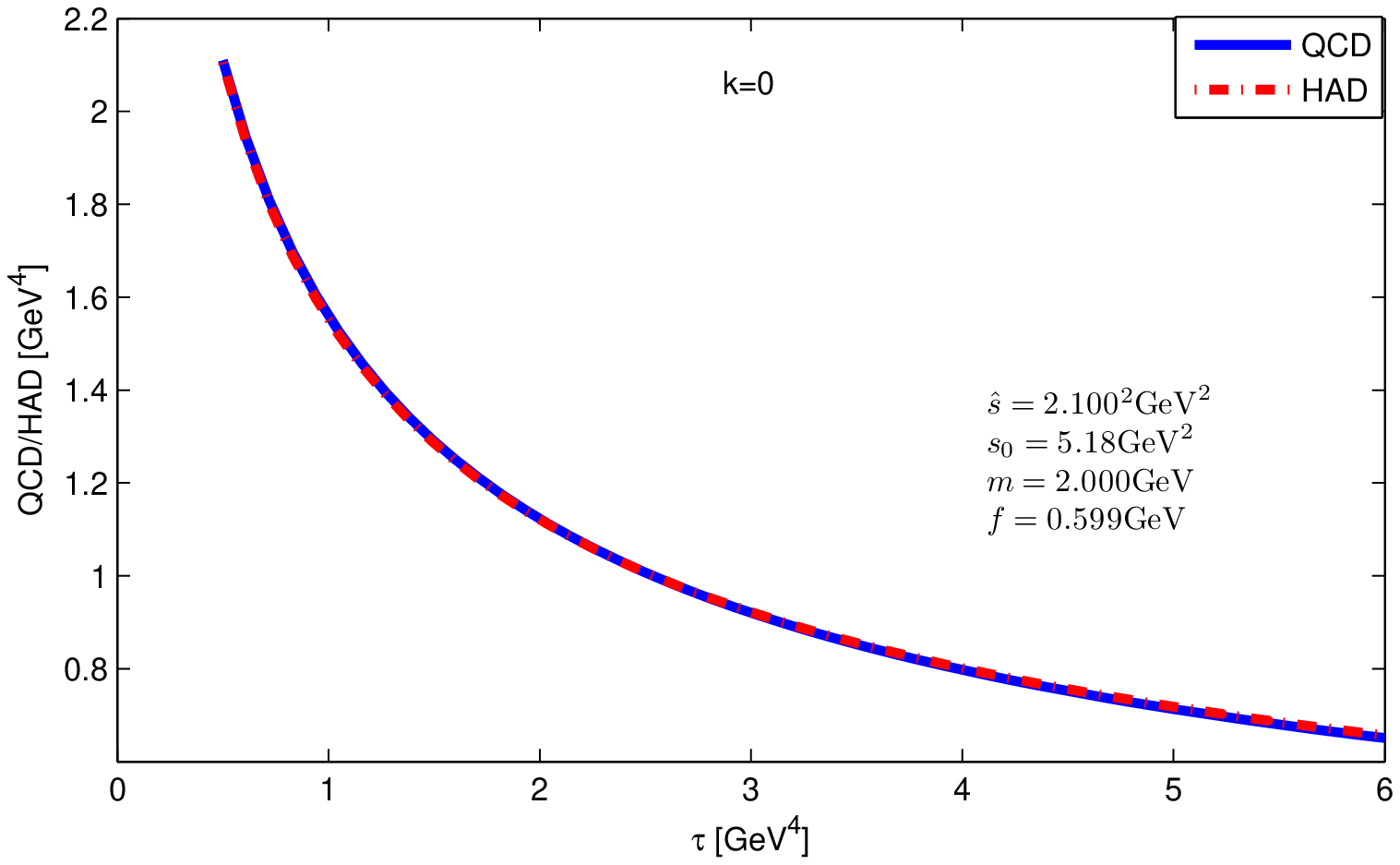}
\includegraphics[width=8.0cm]{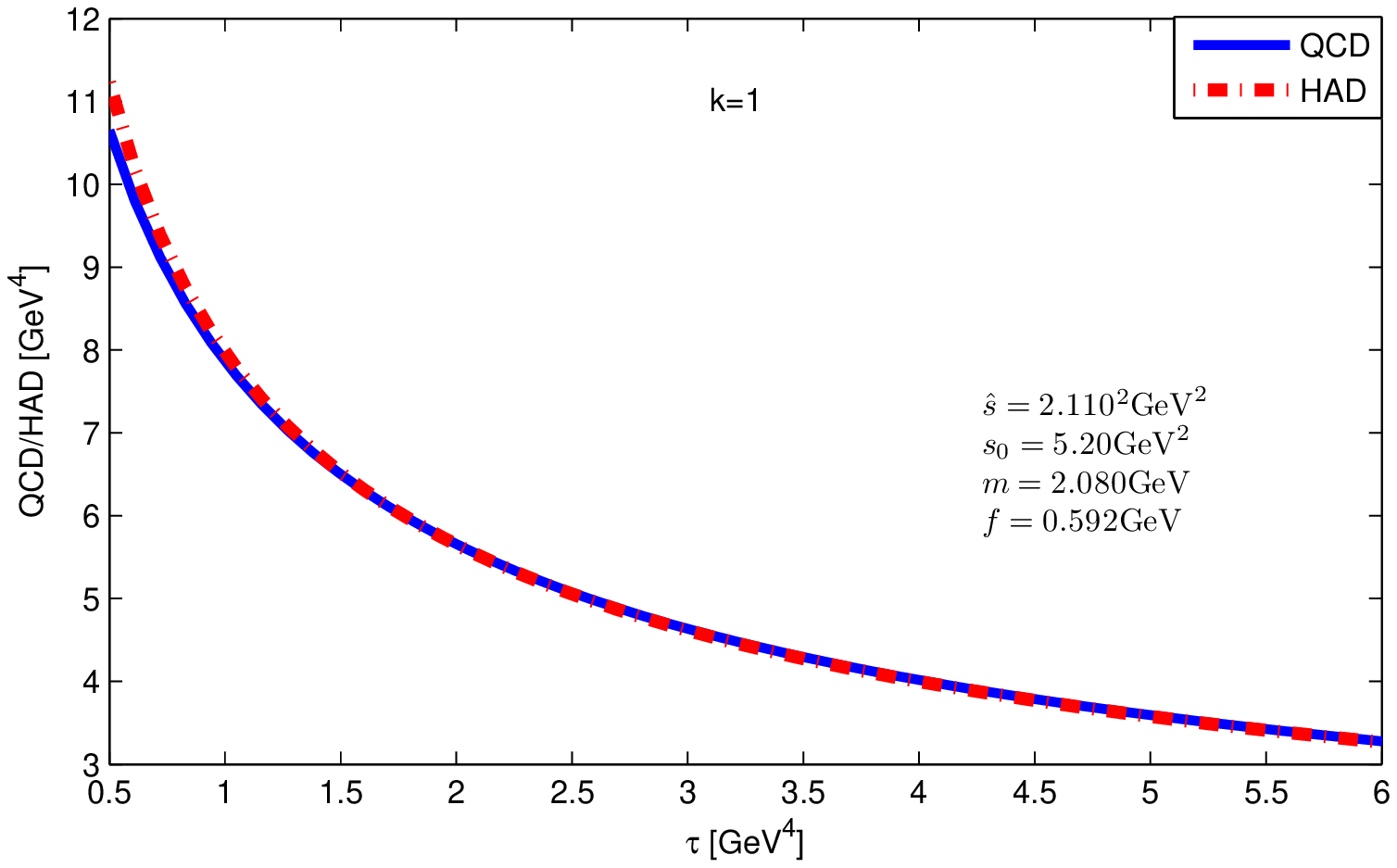}
\caption{\small The lhs (dashed line) and rhs (solid line) of the sum rules \eqref{eq4:42} with $k=-1,0,1$ versus $\tau$ in the case where the correlation function $\Pi^{\textrm{QCD}}(Q^2)$ contains only pure perturbative contribution and pure instanton one, and a single zero-width resonance plus continuum model is adopted for the spectral function.}
\label{Fig:4}
\end{figure}

Let us first consider the case of single-resonance plus continuum model \eqref{eq4:34} of the spectral function by excluding the interference contribution $\Pi^{\textrm{int}}(Q^2)$ from $\Pi^{\textrm{QCD}}(Q^2)$ (the case A), in order to recover the results before. In this case, the imaginary part of $\Pi^{\textrm{QCD}}(Q^2)$, however, becomes negative for $s$ below $3.9\textrm{GeV}^2$, and the full interaction plays a role of repulsive potential in the pseudoscalar channel. This is the reason why the authors in Ref. \cite{NuclPhysA.728.165} cannot find the signal for the pseudoscalar glueball. When we chose to work in the positively definite region, say $s\in(4,10)\textrm{GeV}^2$, The mass of the $0^{-+}$ glueball can be worked out by using the family of Gaussian sum rules \eqref{eq4:42}. The fitting parameters are listed in the first three lines of Tab. \ref{tab:3BWR}  and the corresponding matching curves for $k=-1$, $0$ and $+1$ are displayed in FIG. \ref{Fig:4}, respectively. The optical values of mass, coupling constant and $s_0$ of $0^{-+}$ pseudoscalar glueball are
\begin{eqnarray}
&&m=2.010\pm0.299\textrm{GeV},~f=0.584\pm 0.043\textrm{GeV},\notag\\
&&~ s_0=5.21\pm0.43\textrm{GeV}^2,
\label{eq5:64}
\end{eqnarray}
where the errors are estimated from the uncertainties of the spread between the individual sum rules, and by varying value of $\Lambda$ in the region of (the same for hereafter)
\begin{eqnarray}
\Lambda=120 \sim 200 \textrm{MeV}.
\label{eq5:65}
\end{eqnarray}
The mass values of $0^{-+}$ pseudoscalar glueball in \eqref{eq5:64} are reasonably consist with the one obtained in Ref. \cite{PhysRevD.71.054008} by adding the topological charge screening effect and performing the so-called Gaussian-tail distribution for instanton size. However, after performing the Gaussian-tail distribution for instanton size, the mass of the $0^{++}$ glueball is lower to be around $1.25\textrm{GeV}$ \cite{PhysRevD.71.054008} in contradiction with the lattice simulation \cite{Phys.Rev.D61.014015,PhysRevD.60.034509} and phenomenology \cite{Phys.Lett.B667.1}. The mass scales in \eqref{eq5:64} locate at the strong repulsive potential region of energy (below $3.9\textrm{GeV}^2$) where the bound state of glueball cannot form when working in the spike distribution which is motivated from the liquid instanton model of QCD vacuum in the large $N_c$ limit. In fact, the fundamental spectral positivity bound can be traced back to the definition of the correlation function \eqref{eq2:3}. The spectral function should be positive even before taking the average with any specific instanton-size distribution. It is difficult for us to understand that there is no artificial in changing the positivity behavior of the spectral function just by performing the average with the Gaussian-tail distribution for the instanton size.

\begin{figure}[hbt]
\centering
\includegraphics[width=8.6cm]{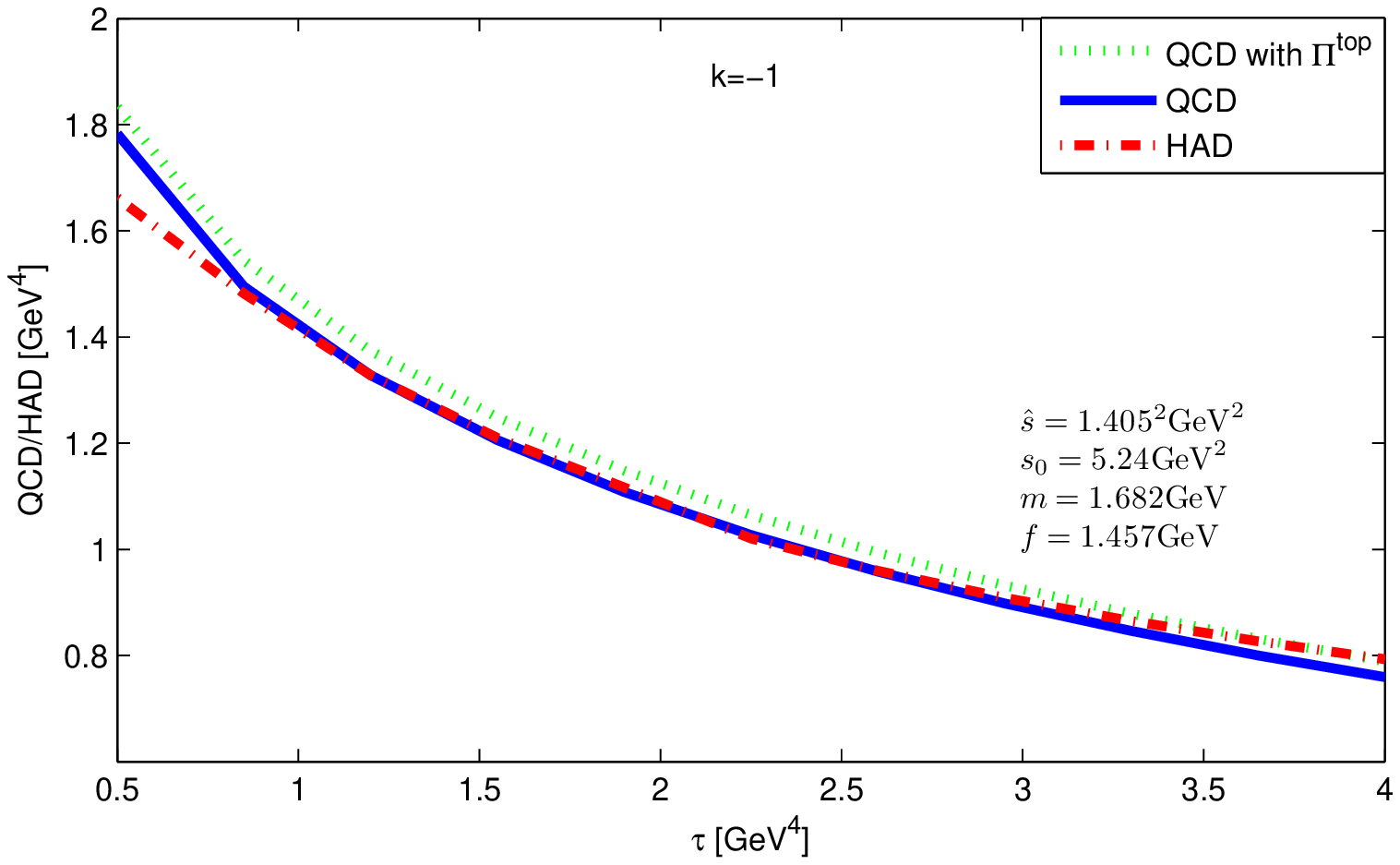}
\includegraphics[width=8.6cm]{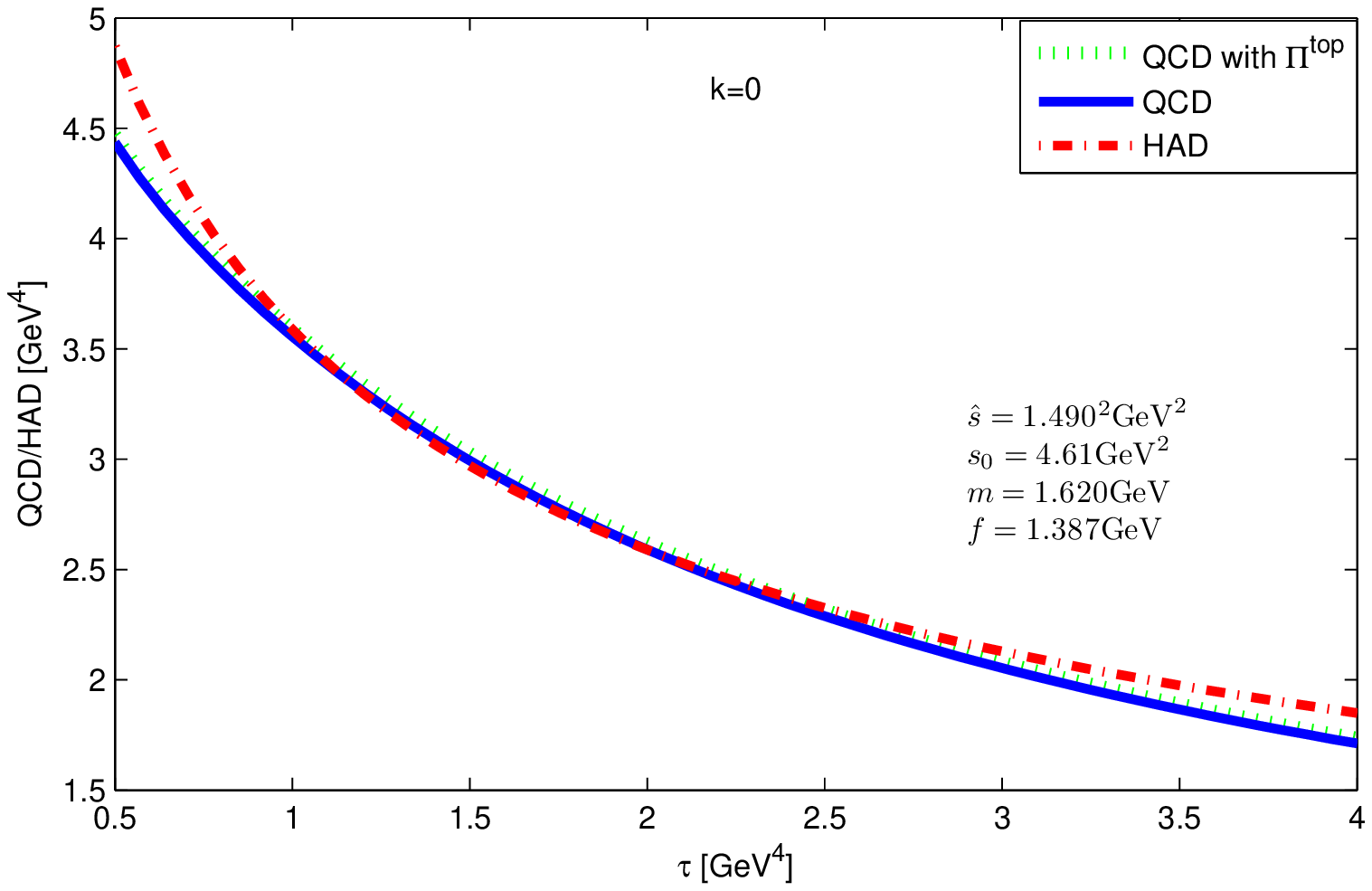}
\includegraphics[width=8.6cm]{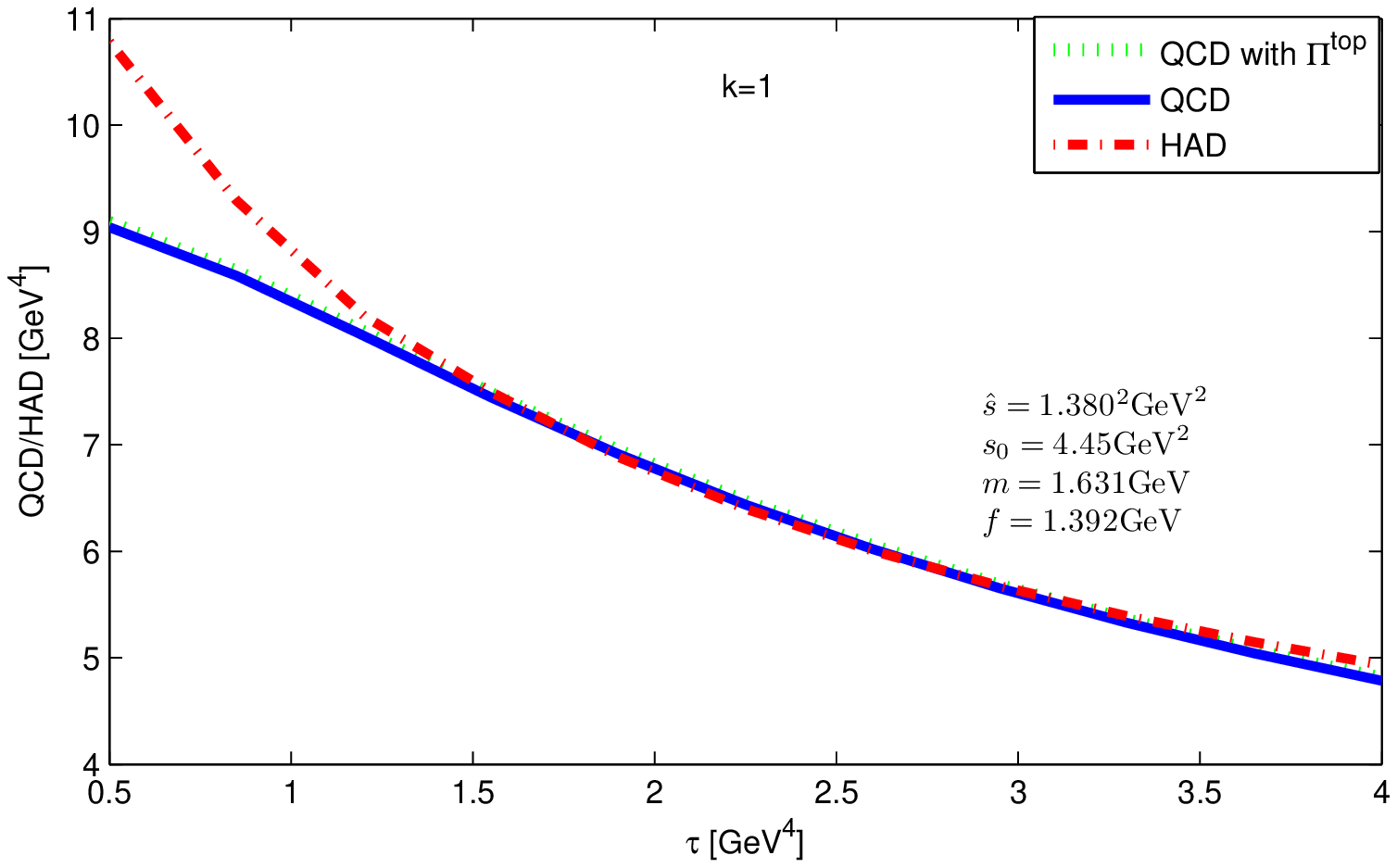}
\caption{\small The lhs without the topological charge screening contribution (dashed line), rhs with the topological charge screening contribution (dot line) and rhs (solid line) of the sum rules \eqref{eq4:42} with $k=-1,0,1$ versus $\tau$ in the case where the interference contribution is included in the correlation function $\Pi^{\textrm{QCD}}(Q^2)$, and a single zero-width resonance plus continuum model is adopted for the spectral function.}
\label{Fig:5}
\end{figure}

\begin{figure}[hbt]
\centering
\includegraphics[width=8.6cm]{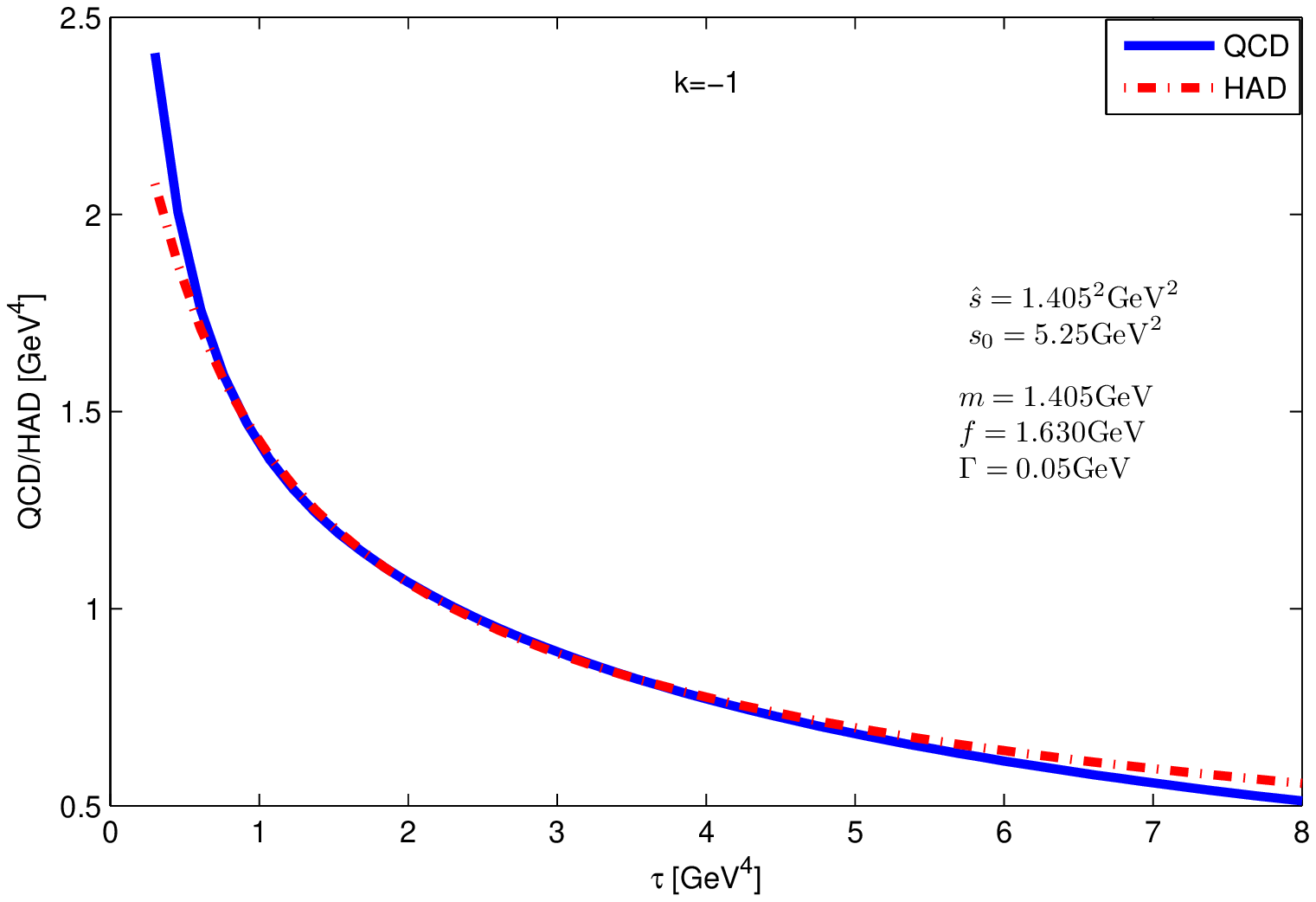}
\includegraphics[width=8.6cm]{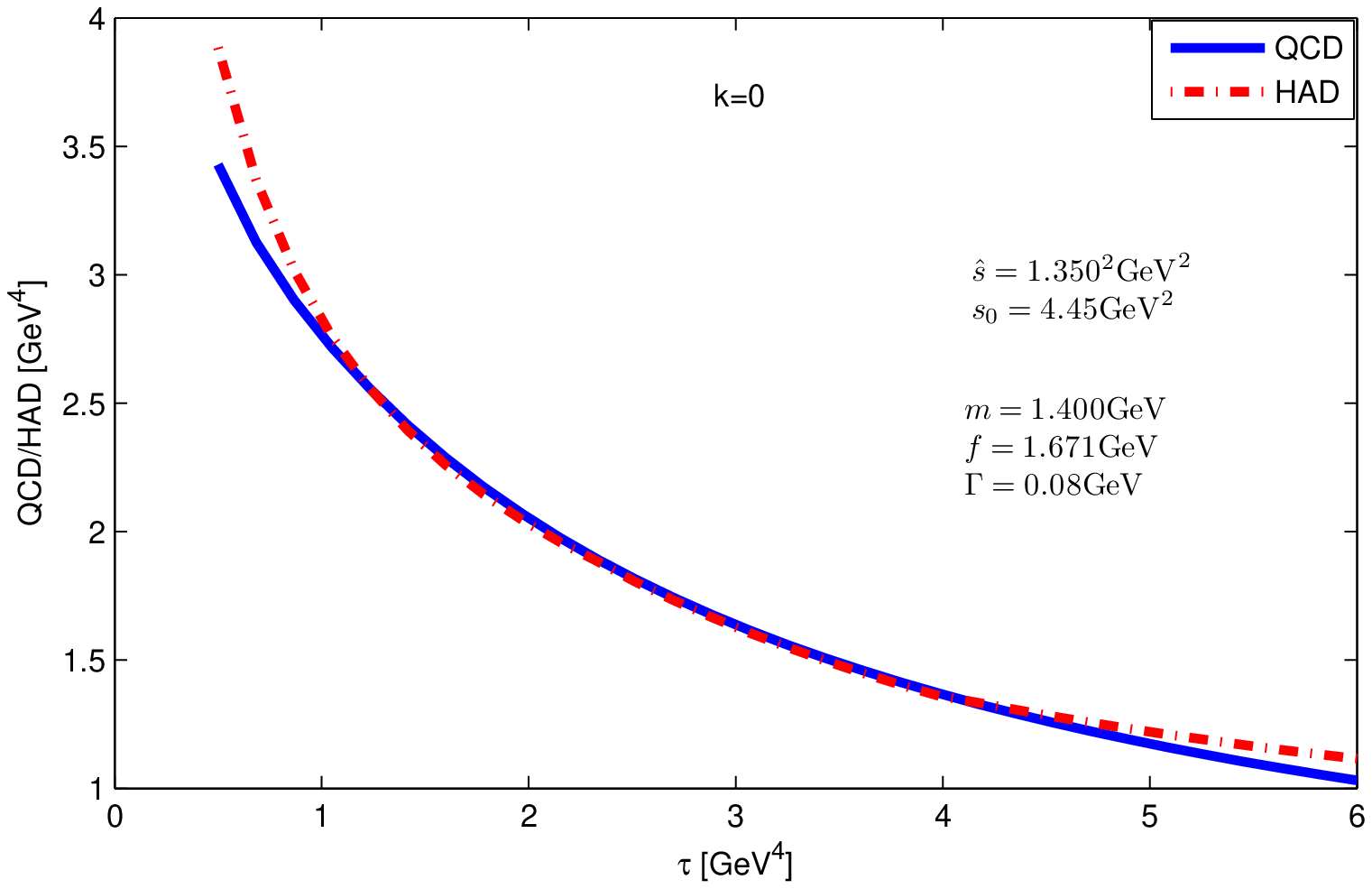}
\includegraphics[width=8.6cm]{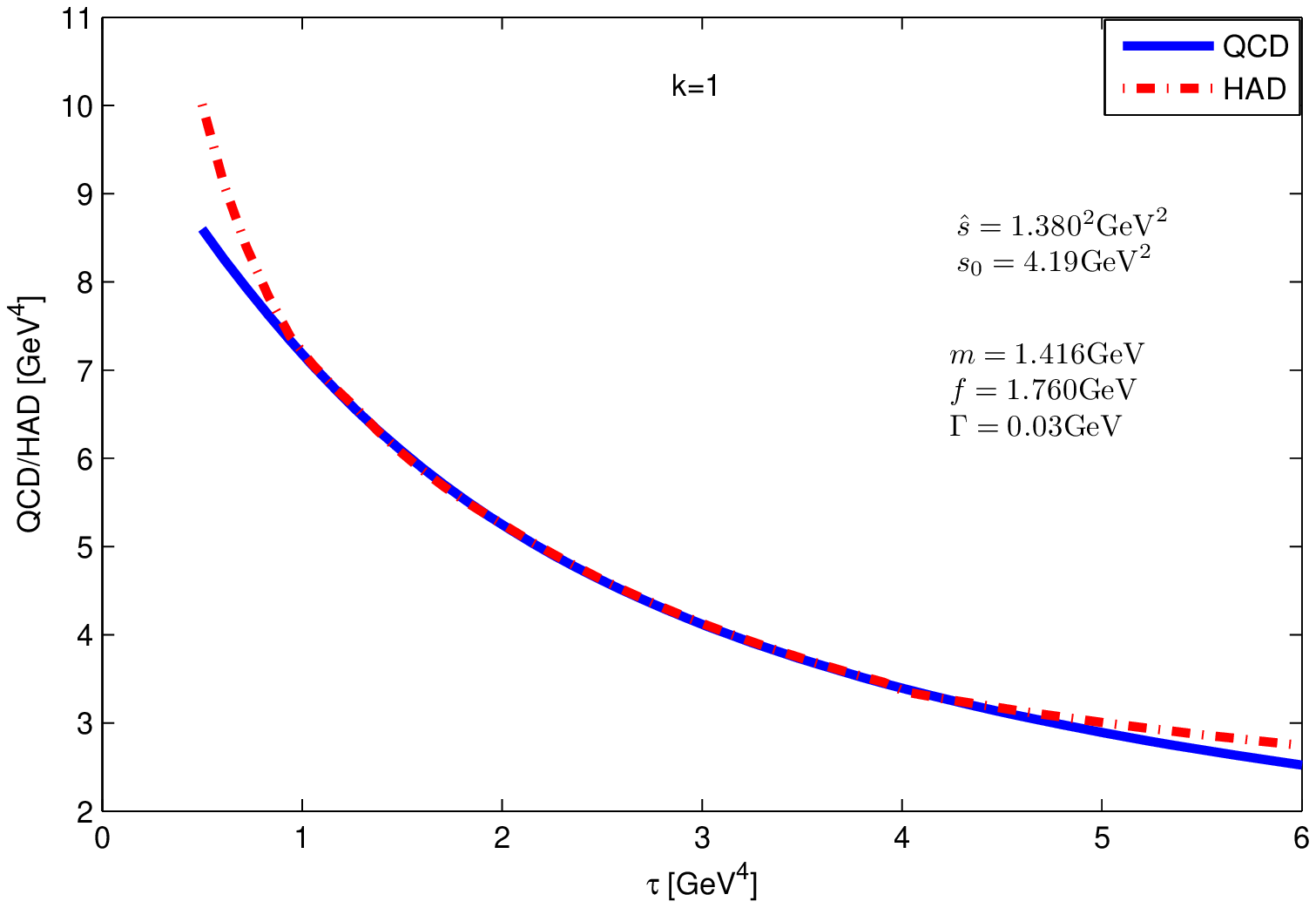}
\caption{\small The lhs (dashed line) and rhs (solid line) of the sum rules \eqref{eq4:42} with $k=-1,0,1$ versus $\tau$ in the case where the correlation function $\Pi^{\textrm{QCD}}(Q^2)$ contains the pure instanton, interference, and pure perturbative contributions, and a single finite-width resonance plus continuum model is adopted for the spectral function.}
\label{Fig:6}
\end{figure}

From now on, the full correlation function $\Pi^{\textrm{QCD}}(Q^2)$ including the interference contribution $\Pi^{\textrm{int}}(Q^2)$, \eqref{eq2:23}, is used in our analysis of the Gaussian sum rules \eqref{eq4:42}. In the case of single-resonance plus continuum models, specified respectively by \eqref{eq4:34} and \eqref{eq4:38}, for the spectral function, the optimal parameters governing the sum rules with zero (the case B) and finite (the case C) widths are listed from the fourth to ninth line of Tab. \ref{tab:3BWR} and the corresponding curves for the lhs and rhs of \eqref{eq4:42} with $k=-1$, $0$ and $+1$ are displayed in FIG. \ref{Fig:5} and \ref{Fig:6} respectively. From Tab. \ref{tab:3BWR}, the optical values of the pseudoscalar glueball mass, width, coupling and the duality threshold with the best matching are:
\begin{eqnarray}
&&m=1.644\pm0.194\textrm{GeV},~f=1.412\pm0.129\textrm{GeV},\notag\\
&&~ s_0=4.77\pm0.74\textrm{GeV}^2,
\label{eq5:66}
\end{eqnarray}
for one zero-width resonance model, and
\begin{eqnarray}
&&m=1.407\pm0.162~\textrm{GeV},~\Gamma=0.053\pm0.018~\textrm{GeV}\notag\\
&&f=1.687\pm0.145~\textrm{GeV},~ s_0=4.63\pm0.62~\textrm{GeV}^2,
\label{eq5:67}
\end{eqnarray}
for one finite-width resonance model. It is shown in FIG. \ref{Fig:5} that the topological charge screening effect has little impact on Gaussian sum rules indeed.

\begin{figure}[hbt]
\centering
\includegraphics[width=8.6cm]{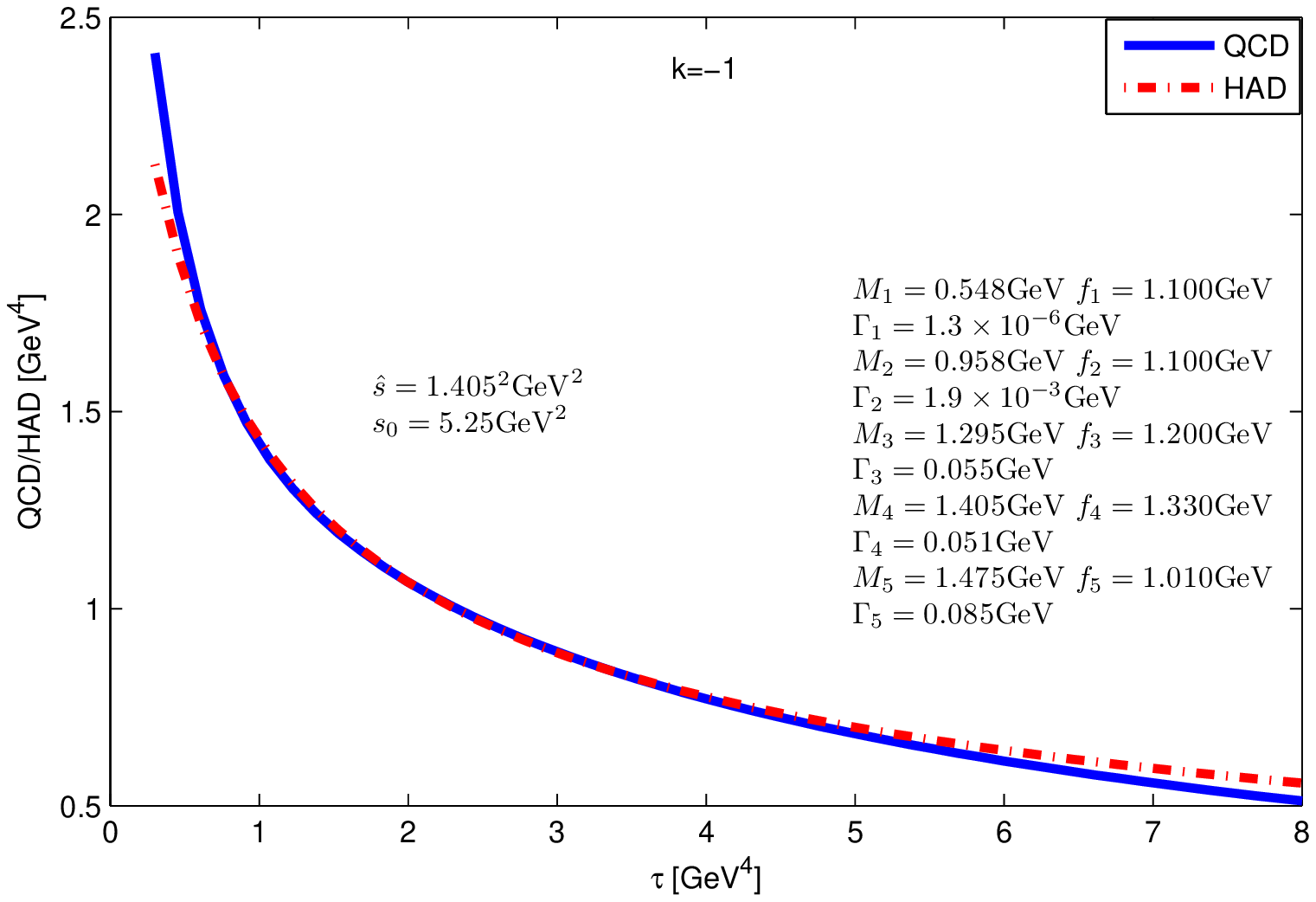}
\includegraphics[width=8.6cm]{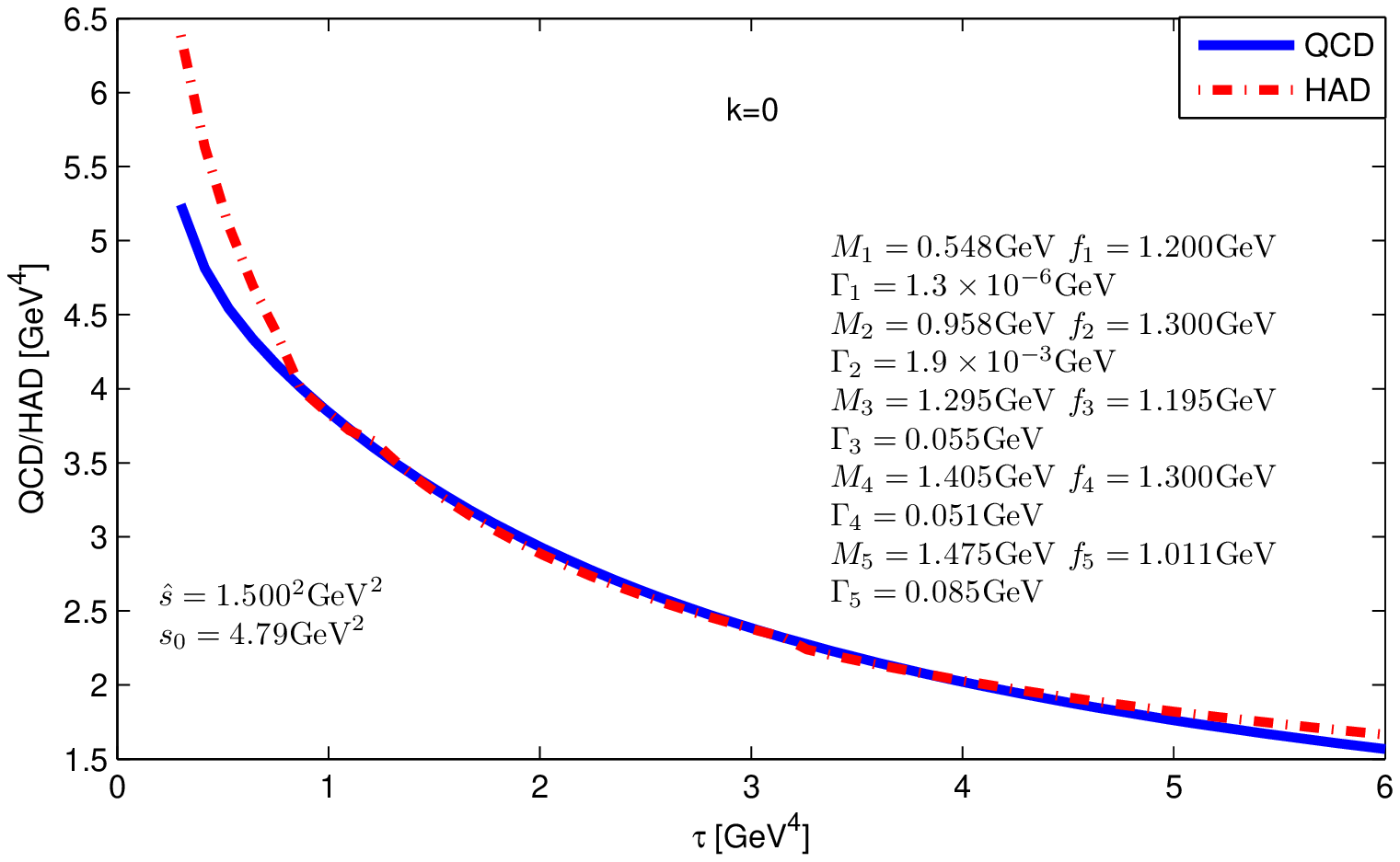}
\includegraphics[width=8.6cm]{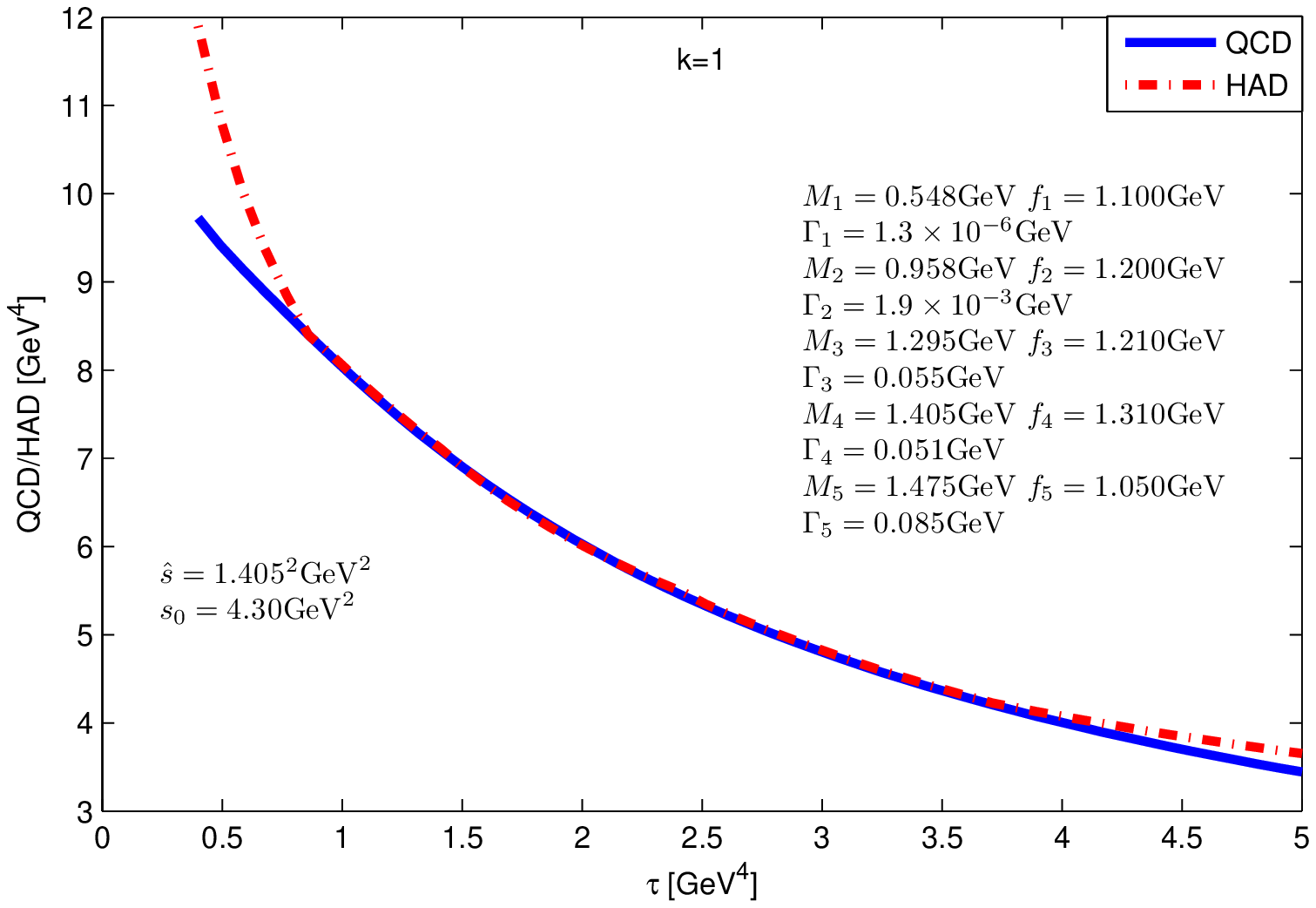}
\caption{\small The lhs (dashed line) and rhs (solid line) of the sum rules \eqref{eq4:42} with $k=-1,0,1$ versus $\tau$ in the case where the correlation function $\Pi^{\textrm{QCD}}(Q^2)$ contains the pure instanton, interference, and pure perturbative contributions, and five finite-width resonance plus continuum model is adopted for the spectral function.}
\label{Fig:7}
\end{figure}

In the numerical simulation for the case of the five finite-width resonances plus continuum model \eqref{eq4:39} for the spectral function (the case D), we just choose the data in PDG as the fitting parameters for masses and width of the resonances $\eta(548)$, $\eta(985)$, $\eta(1295)$ and $\eta(1475)$, and the result of the single resonance model (the case C) as the fitting parameters for $\eta(1405)$; while the couplings of the five resonances to the current are chosen to be approximately the same as that for $\eta(1405)$ determined in case C because the pseudoscalar qurkonia can be directly coupled to the gluon anomaly, and as a consequence, all five resonances should be coupled to the current with the strengths of almost the same magnitude of degree; finally, the optimal parameters are determined by adjusting the chosen parameters so that the matching measure $\delta$ for both sides of the Gaussian sum rules \eqref{eq4:42} is minimal. The optimal parameters governing the sum rules are listed in the remaining lines of Tab. \ref{tab:3BWR}. The corresponding curves for the lhs and rhs of \eqref{eq4:42} with $k=-1$, $0$ and $+1$ are displayed in the FIG. \ref{Fig:7}. Taking the average, the optical values of the widths of the five lowest $0^{-+}$ resonances in the world of QCD with three massless quarks, and the corresponding optical fit parameters are predicted to be
\begin{eqnarray}
&&m_{\eta(548)}=0.548\pm0.022~\textrm{GeV},  f_{\eta(548)}=1.133\pm0.167~\textrm{GeV},\nonumber\\
&&\Gamma_{\eta(548)}=1.3\times10^{-6}\pm3.9\times10^{-8}~\textrm{GeV}
\label{eq5:68}
\end{eqnarray}
\begin{eqnarray}
&&m_{\eta(985)}=0.958\pm0.051~\textrm{GeV},  f_{\eta(985)}=1.200\pm0.233~\textrm{GeV},\nonumber\\
&&\Gamma_{\eta(985)}=1.9\times10^{-3}\pm5.7\times10^{-5}~\textrm{GeV}
\label{eq5:69}
\end{eqnarray}
\begin{eqnarray}
&&m_{\eta(1295)}=1.295\pm0.075~\textrm{GeV},  f_{\eta(1295)}=1.202\pm0.112~\textrm{GeV},\nonumber\\
&&\Gamma_{\eta(1295)}=0.055\pm0.018~\textrm{GeV}
\label{eq5:70}
\end{eqnarray}
\begin{eqnarray}
&&m_{\eta(1405)}=1.405\pm0.081~\textrm{GeV},  f_{\eta(1405)}=1.313\pm0.105~\textrm{GeV},\nonumber\\
&& \Gamma_{\eta(1405)}=0.051\pm0.017~\textrm{GeV}
\label{eq5:71}
\end{eqnarray}
\begin{eqnarray}
&&m_{\eta(1475)}=1.475\pm0.092~\textrm{GeV},  f_{\eta(1475)}=1.023\pm0.097~\textrm{GeV},\nonumber\\
&&\Gamma_{\eta(1475)}=0.085\pm0.028~\textrm{GeV}
\label{eq5:72}
\end{eqnarray}
with
\begin{equation}
s_0=4.78\pm0.64~\textrm{GeV}^2.
\label{eq5:73}
\end{equation}
The FIG. \ref{Fig:6} and \ref{Fig:7} show the satisfactory compatibility between both sides of the sum rules over the whole fiducial region. These results are in good accordance with the experimental discovered resonance \cite{Prog.Part.Nucl.Phys.63.74}
\begin{equation}
m_{\eta(1405)}=1409.8\pm 2.5\textrm{MeV},\,\,\,\,\Gamma_{\eta(1405)}=51.1\pm 3.4\textrm{MeV}
\label{eq5:74}
\end{equation}
\begin{table*}[ht]
 \caption{\small {The optimal fitting values of the mass $m$, width $\Gamma$, coupling constant $f$, continuum threshold $s_0$ , $\hat{s}$, and matching measure $\delta$ for the possible $0^{-+}$ resonances in the sum rule window $[\tau_{\textrm{min}},\tau_{\textrm{max}}]$ for the best matching between lhs and rhs of the sum rules \eqref{eq4:42} with $k=-1,0,1$ are listed, where in case A, the correlation function $\Pi^{\textrm{QCD}}(Q^2)$ contains only pure perturbative contribution and pure instanton one, and a single zero-width resonance plus continuum model is adopted for the spectral function, while all the contributions arising from pure instanton, pure perturbative and interference between both are included in the  correlation function for cases B, C and D, in which a single zero-width resonance plus continuum model of the spectral function is adopted for case B, and a single finite-width resonances plus continuum model for case C, and the five finite-width resonances plus continuum model for case D, respectively.}}
\begin{center}
\begin{tabular}{cc|ccccccccc}\hline\hline
& case& $k$& resonances& $\sqrt{\hat{s}} (\textrm{GeV})$  &$m  (\textrm{GeV})$& $\Gamma  (\textrm{GeV})$& $f  (\textrm{GeV})$& $s_0 (\textrm{GeV}^2)$& $[\tau_{\textrm{min}},\tau_{\textrm{max}}](\textrm{GeV}^4)$& $\delta/10^{-5}$\\
\hline
 &          &$-1$& &$2.050$& $1.950\pm0.050$& 0& 0.561& $5.26\pm0.14$& $[0.8,4.0]$    & $22.60+1.46$\\
 &A         &$ 0$& &$2.100$& $2.000\pm0.035$& 0& 0.599& $5.18\pm0.12$& $[0.5,6.0]$    & $1.99+0.13$\\
  &         &$ 1$& &$2.110$& $2.080\pm0.040$& 0& 0.592& $5.20\pm0.10$& $[0.8,6.0]$    & $7.56+0.45$\\
\hline
 &       &$-1$& &$1.405$& $1.682\pm0.023$& 0& 1.457& $5.24\pm0.15$& $[1.0,3.0]$    & $49.72+2.98$\\
 &B      &$ 0$& &$1.490$& $1.620\pm0.027$& 0& 1.387& $4.61\pm0.09$& $[0.8,2.5]$    & $3.67+0.18$\\
  &      &$ 1$& &$1.380$& $1.631\pm0.031$& 0& 1.392& $4.45\pm0.11$& $[1.4,3.2]$    & $8.97+0.43$\\
\hline
 &       &$-1$& &$1.405$& $1.405\pm0.024$& 0.05& 1.630& $5.25\pm0.14$& $[0.8,4.0]$    & $2.44+0.12$\\
 &C      &$ 0$& &$1.350$& $1.400\pm0.025$& 0.08& 1.671& $4.45\pm0.12$& $[0.5,4.3]$    & $4.47+0.21$\\
  &      &$ 1$& &$1.380$& $1.416\pm0.024$& 0.03& 1.760& $4.19\pm0.11$& $[0.5,4.5]$    & $2.19+0.11$\\
\hline
     & \multirow{18}{*}{D}&   & $\eta(548)$& &$0.548\pm0.008$& $1.3\times10^{-6}$& 1.100& & \\
   &   &    & $\eta(958)$&                   &$0.958\pm0.014$& $1.9\times10^{-3}$& 1.100& & \\
    &   &$-1$& $\eta(1295)$&$1.405$          &$1.295\pm0.020$& 0.055& 1.200& $5.25\pm0.12$& $[0.5,4.5]$& $4.88+0.23$\\
     &  &   & $\eta(1405)$&                  &$1.405\pm0.021$& 0.051& 1.330& &\\
     & &    & $\eta(1475)$&                  &$1.475\pm0.023$& 0.085& 1.010& & \\
\cline{3-11}
     & &    & $\eta(548)$&                   &$0.548\pm0.009$& $1.3\times10^{-6}$& 1.200& & \\
     & &    & $\eta(958)$&                   &$0.958\pm0.016$& $1.9\times10^{-3}$& 1.300& & \\
     &  &$0$& $\eta(1295)$ &$1.500$          &$1.295\pm0.021$& 0.055& 1.195& $4.79\pm0.11$& $[0.5,4.3]$& $3.38+0.17$\\
     &  &   & $\eta(1405)$&                  &$1.405\pm0.022$& 0.051& 1.300& &\\
     & &    & $\eta(1475)$&                  &$1.475\pm0.025$& 0.085& 1.011& & \\
\cline{3-11}
    & &    & $\eta(548)$&                    &$0.548\pm0.010$& $1.3\times10^{-6}$& 1.100& & \\
     & &    & $\eta(958)$&                   &$0.958\pm0.017$& $1.9\times10^{-3}$& 1.200& & \\
     &  &$1$& $\eta(1295)$&$1.405$           &$1.295\pm0.022$& 0.055& 1.210& $4.30\pm0.10$& $[1.0,4.0]$& $3.56+0.17$\\
     &  &   & $\eta(1405)$&                  &$1.405\pm0.023$& 0.051& 1.310& &\\
     & &    & $\eta(1475)$&                  &$1.475\pm0.024$& 0.085& 1.050& & \\
\hline\hline \label{tab:3BWR}
\end{tabular}
\end{center}
\end{table*}

\section{Discussion and Conclusion}
\label{sectionVI}
The instanton-gluon interference and its role in finite-width Gaussian sum rules for $0^{-+}$ pseudoscalar glueball are analyzed in this paper. Our main results can be summarized as follows:

First, the contribution to the correlation function arising from the interference between the classical instanton fields and the quantum gluon ones is reexamined and derived in the framework of the semi-classical expansion of the instanton liquid vacuum model of $\textrm{QCD}$. The resultant expression is gauge invariant, and free of the infrared divergence, and differs from our previous one not only in some coefficients but also in the logarithms structures\cite{Eur.Phys.J.Plus128.115,J.Phys.G.41.035004}. Its magnitude is just between the larger contribution from pure classical instanton configurations and the smaller one from the pure quantum fields, and plays a great role in sum rule analysis in accordance with the spirit of semi-classical expansion. The imaginary part of the correlation function including this interference contribution turns to be positive without including of the topological charge screening effect which is proven to be smaller than the perturbative contribution in the fiducial sum rule window, and negligible in comparison with the interference effect. The so-called problem of positivity violations in the imaginary part of the correlation function, stressed by H. Forkel\cite{PhysRevD.71.054008}, disappears. Moreover, it is excluded in the correlation function the traditional condensate contribution to avoid the double counting\cite{PhysRevD.71.054008} because condensates can be reproduced by the instanton distributions\cite{NuclPhysB.203.116,Phys.Lett.B147.351,PhysRevLett.49.259,J.Phys.G21.751}; Another cause to do so is that the usual condensate contribution is proven to be unusually weak, and cannot fully reflect the nonperturbative nature of the low-lying gluonia\cite{NuclearPhysicsB.165.55,PhysRevD.71.054008,Eur.Phys.J.Plus128.115,J.Phys.G.41.035004}; In our opinion, the condensate contribution can be considered as a small fraction of the corresponding instanton one, so it is naturally taken into account already.

Second, the properties of the lowest lying $0^{-+}$ pseudoscalar glueball are systematically investigated in a family of Gaussian sum rules in five different cases. In case A, the correlation function $\Pi^{\textrm{QCD}}(Q^2)$ contains only pure perturbative contribution and pure instanton one, and a single zero-width resonance plus continuum model is adopted for the spectral function, and of course, the old results are recovered (even excluding the topological charge screening contribution), and some pathology is explored. To go beyond the above constraint, all the contributions arising from pure instanton, pure perturbative and interference between both are included in the correlation function for cases B, C and D, in which a single zero-width resonance plus continuum model of the spectral function is adopted for case B, and a single finite-width resonances plus continuum model for case C, and the five finite-width resonances plus continuum model for case D, respectively. The optimal fitting values of the mass $m$, width $\Gamma$, coupling constant $f$, continuum threshold $s_0$ for the possible $0^{-+}$ resonances are obtained, and quite consistent with each other. The main difference between this work and our previous one\cite{Eur.Phys.J.Plus128.115,J.Phys.G.41.035004} is that for the spectral function of considered resonances, instead of the zero-width approximation of one gluonic resonance plus another low-lying quark-antiquartk ones, the finite-width Breit-Wigner form with a correct threshold behavior for the lowest five resonances with the same quantum numbers is used in this work in order to compare with the phenomenology. The resultant Gaussian sum rules with $k=-1,0,+1$ are carried out with a few of the $\textrm{QCD}$ standard inputting parameters, and really in accordance with the experimental data.

As a discussion, let us now identify where the lowest lying $0^{-+}$ pseudoscalar glueball is. The result of the single-resonance plus continuum models B and C, namely Eqs. \eqref{eq5:66} and \eqref{eq5:67}, imply that the meson $\eta(1405)$ may be the most fevered candidate for the lowest lying $0^{-+}$ pseudoscalar glueball because the difference between the two models is just the width of the resonances, and the latter is of course believed to be more in accordance with the reality. This conclusion can further be justified by the result of the five-resonances plus continuum model, namely Eqs. \eqref{eq5:71}, \eqref{eq5:72} and \eqref{eq5:73}. Note that the first two resonances $\eta(548)$ and $\eta(985)$ are far away from the mass scale of $\eta(1405)$, and usually considered as the superposition of the fundamental flavor-singlet and octet pseudoscalar mesons composed of quark-antiquark pair to have a dominant role in responsible for the axial anomaly\cite{PhysRevD.58.114006,PhysicsLetterB.449.339,Phys.Rev.D90.074019}. In order to explore the structures of the remainder three resonance $\eta(1295)$, $\eta(1405)$ and $\eta(1475)$, we would like use the $\eta$-$\eta'$-$G$ mixing formalism based on the anomalous Ward identity for transition matrix elements\cite{Phys.Lett.B648.267,PhysRevD.79.014024} to relate the physical states $\eta$, $\eta'$ and $G$ to the fundamental flavor-singlet and octet quark-antiquark mesons and the lowest pure gluon state through a rotation
\begin{equation}
\begin{pmatrix}\langle{0}|\hat{O}_p|\eta_{1295}\rangle  \\\langle{0}|\hat{O}_p|\eta_{1475}\rangle \\ \langle{0}|\hat{O}_p|\eta_{1405}\rangle \end{pmatrix}=U\begin{pmatrix}\langle{0}|\hat{O}_p|\eta_8\rangle \\ \langle{0}|\hat{O}_p|\eta_1\rangle \\ \langle{0}|\hat{O}_p|G\rangle \end{pmatrix},
\label{eq6:75}
\end{equation}
where the $U$ is the mixing matrix \cite{PhysRevD.79.014024,Phys.Lett.B648.267}
\begin{equation}
U=\left(
\begin{array}{ccc}
 \cos \text{$\varphi_p $} & -\sin \text{$\varphi_p $} & 0 \\
 \cos \text{$\phi_G $} \sin \text{$\varphi_p $} & \cos \text{$\varphi_p $} \sin \text{$\phi_G $} & \sin \text{$\phi_G $} \\
 -\cos \text{$\phi_G $} \sin \text{$\varphi_p $} & -\cos \text{$\varphi_p $} \sin \text{$\phi_G $} & \cos \text{$\phi_G $}
\end{array}
\right)
\label{eq6:76}
\end{equation}
where $\varphi_p\approx 40^{\circ}$ and $\phi_G \approx 22^{\circ}$ is $\eta-\eta'$ \cite{Phys.Lett.B648.267} mixing angle and the mixing angle of the pseudoscalar glueball with $\eta'$. Then, one has
\begin{equation}
U=\left(
\begin{array}{ccc}
 0.766044 & -0.642788 & 0 \\
 0.595982 & 0.286965 & 0.374607 \\
 -0.595982 & -0.286965 & 0.927184
\end{array}
\label{eq6:77}
\right)
\end{equation}
To be quantitative, the corresponding normalized couplings $F$ to the three resonances $\eta$, $\eta'$ and G with masses 1.295, 1.475, 1.405 $\textrm{GeV}$ can be read from Tab. \ref{tab:3BWR} to be
\begin{eqnarray}
0.51\textrm{GeV}^3, ~0.65\textrm{GeV}^3, ~0.56\textrm{GeV}^3,
\label{eq6:78}
\end{eqnarray}
after normalization, respectively. As a relatively rough estimation, from Eq. \eqref{eq6:78} the values of the couplings of $\hat{O}_p$ to the states $\eta_1$, $\eta_8$ and pseudoscalar glueball $G$ are obtained
\begin{eqnarray}
\langle{0}|\hat{O}_p|\eta_1\rangle&=&-0.117\textrm{GeV}^3,\notag\\
\langle{0}|\hat{O}_p|\eta_8\rangle&=&0.568\textrm{GeV}^3,\notag\\
\langle{0}|\hat{O}_p|G\rangle&=&0.813\textrm{GeV}^3.
\label{eq6:79}
\end{eqnarray}
by reversing Eq. \eqref{eq6:75}. This shows that the coupling of the $0^{-+}$ glueball current to the pure glueball state $G$ is dominant, and the signs of the couplings $\langle{0}|\hat{O}_p|\eta_1\rangle$ and $\langle{0}|\hat{O}_p|\eta_8\rangle$ are similar to those predicted by the scalar glueball-meson coupling theorems \cite{NuclPhysB.165.67,NuclPhysB.191.301}.

Furthermore, $\eta(1405)$ and $\eta(1475)$ could originate from a single pole (see Amsler and Masoni's review for eta(1405) in PDG). To check whether the single pseudoscalar meson assumption may be consistent with our sum rule approach, or the dependence of the results on the model selected for the spectral function, we add an analysis of four resonance model for the spectral function. The result is shown in Appendix F. From FIG. \ref{Fig:10} and Tab. 2 where it is easy to see that the four resonance model for the spectral function is inconsistent with our sum rule analysis because, firstly the resultant mass scale of the fourth resonance is approximately $1.41\mathrm{GeV}$ in average which locates outsides of the sum rule window; secondly the match degree $\delta$ of the four resonance model is obviously worse than the $\delta$ of the five resonance model.

In summary, our result suggests that $\eta(1405)$ is a good candidate for the lowest $0^{-+}$ pseudoscalar glueball with some mixture with the nearby excited isovector and isoscalar $q\bar{q}$ mesons. This is a first theoretical support for the phenomenological estimation from the sum rule approach.

\section*{Acknowledgements}
This work is supported by BEPC National Laboratory Project R\&D and BES Collaboration Research Foundation.

\appendix
\section{Pure instanton and perturbative contribution}
\label{A}
 In instanton liquid model, the pure instanton contribution of $0^{-+}$ pseudoscalar glueball with spike instanton distribution in Euclidean space-time is known as\cite{PhysRevD.71.054008,NuclPhysA.686.393,Phys.Atom.Nucl.63.1448,NuclPhysB203.93}
 \begin{equation}
\Pi_E^{(\textrm{cl})}(Q^2)=-2^5\pi^2\bar{n}\bar{\rho}^4Q^4K^2_2(Q\bar{\rho}).
\label{eqA:1}
\end{equation}
where $K^2_2$ is the McDonald functions and $n(\rho)$ is the size distribution of instantons, and the appearance of the overall minus sign in rhs of \eqref{eqA:1} is due to the anti-hermitian property of the current, i.e. $O^{\dag}_{p}=-O_{p}$ in Euclidean space-time.

The second part of \eqref{eq2:10} has already been calculated up to three-loop level in the $\overline{\textrm{MS}}$ dimensional regularization scheme\cite{NuclPhysA.695.205,NuclPhysA.686.393,PhysRevLett.79.353}
\begin{eqnarray}
\Pi^{(\textrm{qu})}(Q^2)&=&\left(\frac{\alpha_s}{\pi}\right)^2Q^4\ln\frac{Q^2}{\mu^2}
\left[a_0+a_1\ln\frac{Q^2}{\mu^2}\right.\notag\\
&+&\left.a_2\ln^2\frac{Q^2}{\mu^2}\right],
\label{eqA:2}
\end{eqnarray}
where $\mu$ is the renormalization scale, and the coefficients with the inclusion of the correct threshold effect are
\begin{eqnarray}
a_0&=&-2\left[1+20.75
\left(\frac{\alpha_s}{\pi}\right)+305.95\left(\frac{\alpha_s}{\pi}\right)^2\right]
,\nonumber\\
a_1&=&2\left(\frac{\alpha_s}{\pi}\right)\left[\frac9{4}
+72.531\left(\frac{\alpha_s}{\pi}\right)\right],\nonumber \\
a_2&=&-10.1250\left(\frac{\alpha_s}{\pi}\right)^{2}
\label{eqA:3}
\end{eqnarray}
for QCD with three massless quark flavors up to three-loop level.

\section{Topological charge screening}
\label{B}
The topological charge screening effect can be understood by tracing back to the anomaly of the axial-vector current $J^5_{\mu}=\Sigma_{i}\bar{\psi}_i\gamma_5\gamma_{\mu}\psi_i$ with $\psi_i$ being quark field of flavor $i$
\begin{eqnarray}
\partial_{\mu}J^5_{\mu}=2N_f Q(x)
\label{eqB:1}
\end{eqnarray}
where the topology charge current $Q(x)$ relates to the pseudoscalar glueball current as $Q(x)=O_p(x)/(8\pi)$. Eq. \eqref{eqB:1} indicates that instantons generate quark-antiquark pairs, which then may form a light meson to mediate the long-range multi-instanton interaction \cite{RevModPhys.70.323}. Kikuchi and Wudka \cite{Phys.Lett.B284} suggest that such a light meson, which can couple to the instanton in the way that the instanton generates all light quark flavors with the identical probability, be the meson $\eta_0$, and constructs an effective Lagrangian (see also \cite{Phys.Lett.B285,Phys.Rev.D52.295}) which gives rise to the topological charge screening contribution $\Pi^{\textrm{top}}$ to the correlation function of the pseudoscalar current as\cite{PhysRevD.71.054008}
\begin{eqnarray}
\Pi^{\textrm{top}}(Q^2)=\frac{F^2_{\eta}}{Q^2+m^2_{\eta}}+\frac{F^2_{\eta'}}{Q^2+m^2_{\eta'}},
\label{eqB:2}
\end{eqnarray}
where $m_{\eta}$ and $m_{\eta'}$ are the masses of the mesons $\eta$ and $\eta'$, and the two constants $F_{\eta}$ and $F_{\eta'}$ are evaluated to be $16\pi\bar{n}\xi\sin\phi$ and $16\pi\bar{n}\xi\cos\phi$, and $\xi$ and $\phi\approx 22^\circ$ are the coupling strength of $\eta_0$ to an instanton and the $\eta-\eta'$ mixing angle, respectively.

\section{The operators $O_i(x)$}
\label{C}
The operators $O_i$ in terms of instanton and quantum gluon fields are
\begin{eqnarray}
O_1(x)&=&F_{\mu\nu,a}[A(x)]F_{\rho\sigma,a}[A(x)]\nonumber\\
O_2(x)&=&4F_{\mu\nu,a}[A(x)](\partial_{\rho}a_{\sigma a}[A(x)])\nonumber\\
O_3(x)&=&4g_s f_{abc} F_{\mu\nu,a}[A(x)]A_{\rho b}(x)a_{\sigma c}(x)\nonumber\\
O_4(x)&=&4(\partial_{\mu}a_{\nu a}(x))(\partial_{\rho}a_{\sigma a}(x))\nonumber\\
O_5(x)&=&8g_sf_{abc}A_{\rho b}(x)(\partial_{\mu}a_{\nu a}(x))a_{\sigma c}(x)\nonumber\\
O_6(x)&=&4g_s^2f_{abc}f_{ade}A_{\mu b}(x)A_{\rho d}(x) a_{\nu c}(x)a_{\sigma e}(x)\nonumber\\
O_7(x)&=&2g_s f_{abc}F_{\mu\nu,a}[A(x)]a_{\rho b}(x)a_{\sigma c}(x)\nonumber\\
O_8(x)&=&4g_s f_{abc}a_{\mu b}(x)a_{\nu c}(x)(\partial_{\rho}a_{\sigma a}(x))\nonumber\\
O_9(x)&=&4g_s^2 f_{abc}f_{ade}A_{\rho d}(x)a_{\mu b}(x)a_{\nu c}(x)a_{\sigma e}(x)\nonumber\\
O_{10}(x)&=&g_s^2 f_{abc}f_{ade}a_{\mu b}(x)a_{\nu c}(x)a_{\rho d}(x)a_{\sigma e}(x)
\label{C1}
\end{eqnarray}
where $F_{\mu\nu,a}[A(x)]$ is the instanton field strength associated with the instanton field $A$.

\section{The interference contributions $\Pi_i^{(\textrm{int})}(q^2)$}
\label{D}
The expressions of the interference contributions $\Pi_i^{(\textrm{int})}(q^2)$ in terms of $O_i$ are
\begin{eqnarray}
&\Pi^{(\textrm{cl}+\textrm{qu})}_1(q^2)&=\hat{T} \langle\Omega|O_2(x)O_2(0)|\Omega\rangle \notag \\
&\Pi^{(\textrm{cl}+\textrm{qu})}_2(q^2)&=\hat{T} \langle\Omega|O_2(x)O_3(0)|\Omega\rangle \notag\\
&\Pi^{(\textrm{cl}+\textrm{qu})}_3(q^2)&=\hat{T} \langle\Omega|O_3(x)O_3(0)|\Omega\rangle \notag\\
&\Pi^{(\textrm{cl}+\textrm{qu})}_4(q^2)&=\hat{T} \langle\Omega|O_4(x)O_5(0)|\Omega\rangle \notag\\
&\Pi^{(\textrm{cl}+\textrm{qu})}_5(q^2)&=\hat{T} \langle\Omega|O_4(x)O_6(0)|\Omega\rangle \notag\\
&\Pi^{(\textrm{cl}+\textrm{qu})}_6(q^2)&=\hat{T} \langle\Omega|O_4(x)O_7(0)|\Omega\rangle \notag\\
&\Pi^{(\textrm{cl}+\textrm{qu})}_7(q^2)&=\hat{T} \langle\Omega|O_5(x)O_5(0)|\Omega\rangle \notag\\
&\Pi^{(\textrm{cl}+\textrm{qu})}_8(q^2)&=\hat{T} \langle\Omega|O_7(x)O_7(0)|\Omega\rangle \notag\\
&\Pi^{(\textrm{cl}+\textrm{qu})}_9(q^2)&=\hat{T} \langle\Omega|O_5(x)O_7(0)|\Omega\rangle \notag\\
&\Pi^{(\textrm{cl}+\textrm{qu})}_{10}(q^2)&=\hat{T} \langle\Omega|O_6(x)O_7(0)|\Omega\rangle \notag\\
&\Pi^{(\textrm{cl}+\textrm{qu})}_{11}(q^2)&=\hat{T} \langle\Omega|O_5(x)O_6(0)|\Omega\rangle \notag\\
&\Pi^{(\textrm{cl}+\textrm{qu})}_{12}(q^2)&=\hat{T} \langle\Omega|O_6(x)O_6(0)|\Omega\rangle
\label{D1}
\end{eqnarray}
where
\begin{align}
\hat{T}\equiv &-\frac{1}{2}\alpha^2_s\bar{n}\epsilon_{\mu\nu\rho\sigma}
\epsilon_{\mu'\nu'\rho'\sigma'}\int{d^4z} \int d^4x.
\label{D2}
\end{align}

\section{Comparison between the condensate contributions and the instanton-induced ones}
\label{E}
\begin{figure}[!hbt]
\centering
\includegraphics[width=8.6cm]{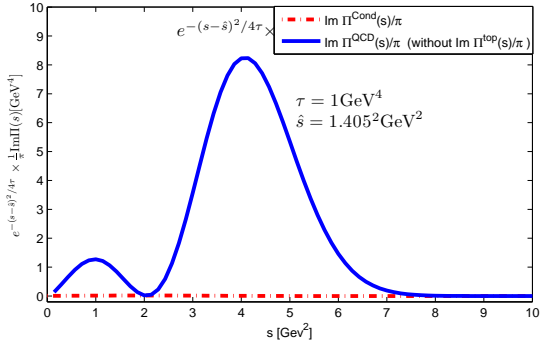}
\caption{\small The contributions to the imaginary part of the correlation function from the condensate (dashed-dotted line) and the total contribution (solid line) versus $s$.}
\label{Fig:8}
\end{figure}

\begin{figure}[!hbt]
\centering
\includegraphics[width=8.6cm]{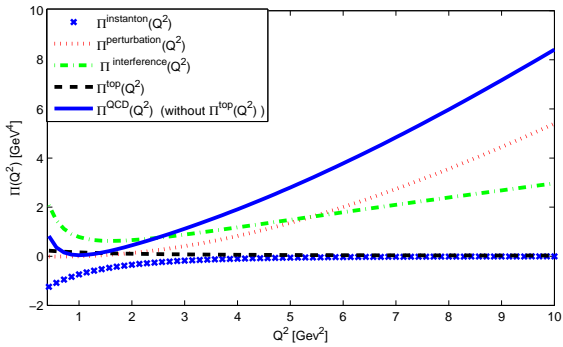}
\caption{\small The contributions to the correlation function arising from the pure instanton (cross line), interference (dashed-dotted line), pure perturbative (dotted line) and topological charge screening (dashed line) and the total contribution without the topological charge screening one (solid line) versus $Q^2$.}
\label{Fig:9}
\end{figure}

The contributions arising from condensates up to the eighth dimensions are known to be as follows \cite{PhysRevD.71.054008}
\begin{eqnarray}
\Pi^{\textrm{cond}}(Q^2)&=&4\alpha_s\langle\alpha_sG^2\rangle+\frac{9}{\pi}\alpha^2_s\langle\alpha_sG^2\rangle\ln{\frac{Q^2}{\mu^2}}\notag\\
&-&8\alpha^2_s\langle g G^3\rangle\frac{1}{Q^2}+\frac{15\pi}{2}\alpha^2_s\langle\alpha_sG^2\rangle^2\frac{1}{Q^4},
             \label{E1}
\end{eqnarray}
where
\begin{eqnarray}
\langle\alpha_sG^2\rangle&=&0.05 \textrm{GeV}^4\notag\\
 \langle gG^3\rangle&=&0.27\textrm{GeV}^2\langle\alpha_sG^2\rangle,
             \label{E2}
\end{eqnarray}
The imaginary part of condensates \eqref{E1} has the form
\begin{eqnarray}
\frac{1}{\pi}\textrm{Im}\Pi^{\textrm{Cond}}(s)&=&-\frac{9}{\pi}\alpha^2_s\langle\alpha_sG^2\rangle-8\alpha^2_s\langle g G^3\rangle\delta(s)\notag\\
&&+\frac{15\pi}{2}\alpha^2_s\langle\alpha_sG^2\rangle^2\delta'(s).
\label{E3}
\end{eqnarray}
The comparison between the imaginary part of the correlation function, \eqref{eq2:24}, and condensate contribution to it are shown in FIG. \ref{Fig:8}, while the comparison between the various real parts of \eqref{eq2:23} (which altogether are related with the imaginary part by the dispersion relation \eqref{eq4:40}) and condensate contribution to it are shown in FIG. \ref{Fig:9}.

\section{The results for the four finite-width resonance model from the Gaussian sum rules}
\label{F}
\begin{figure}[hbt]
\centering
\includegraphics[width=8.6cm]{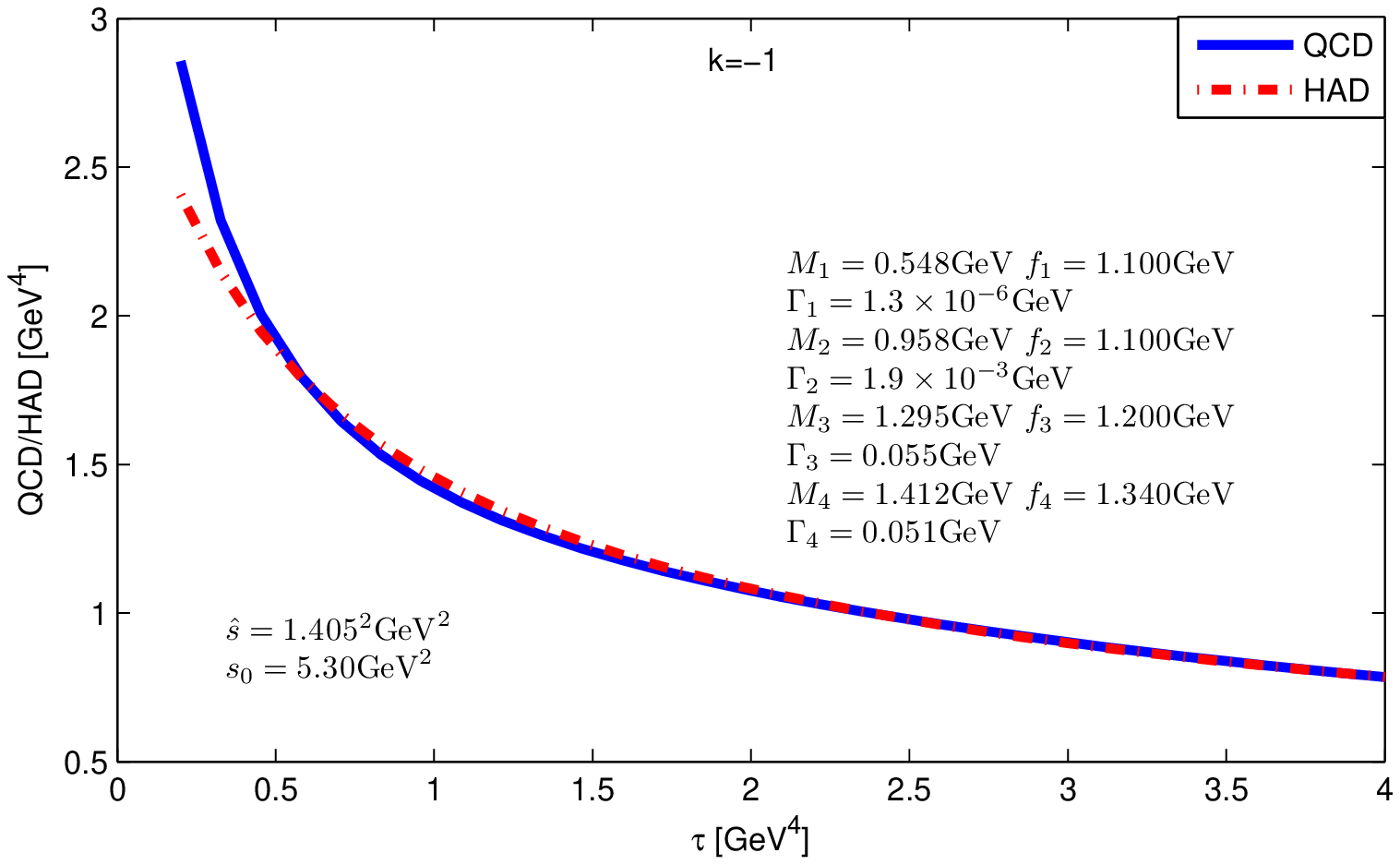}
\includegraphics[width=8.6cm]{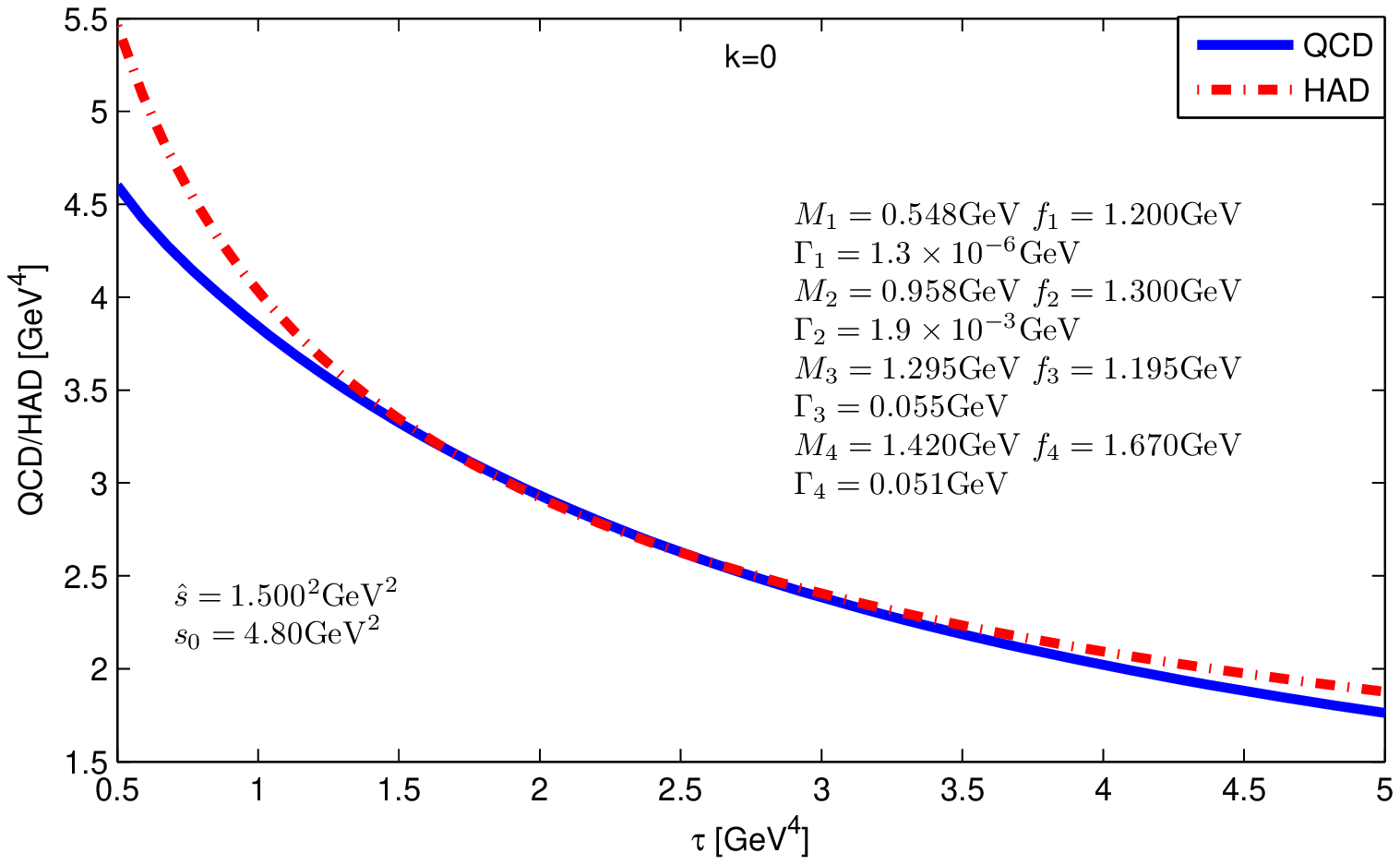}
\includegraphics[width=8.6cm]{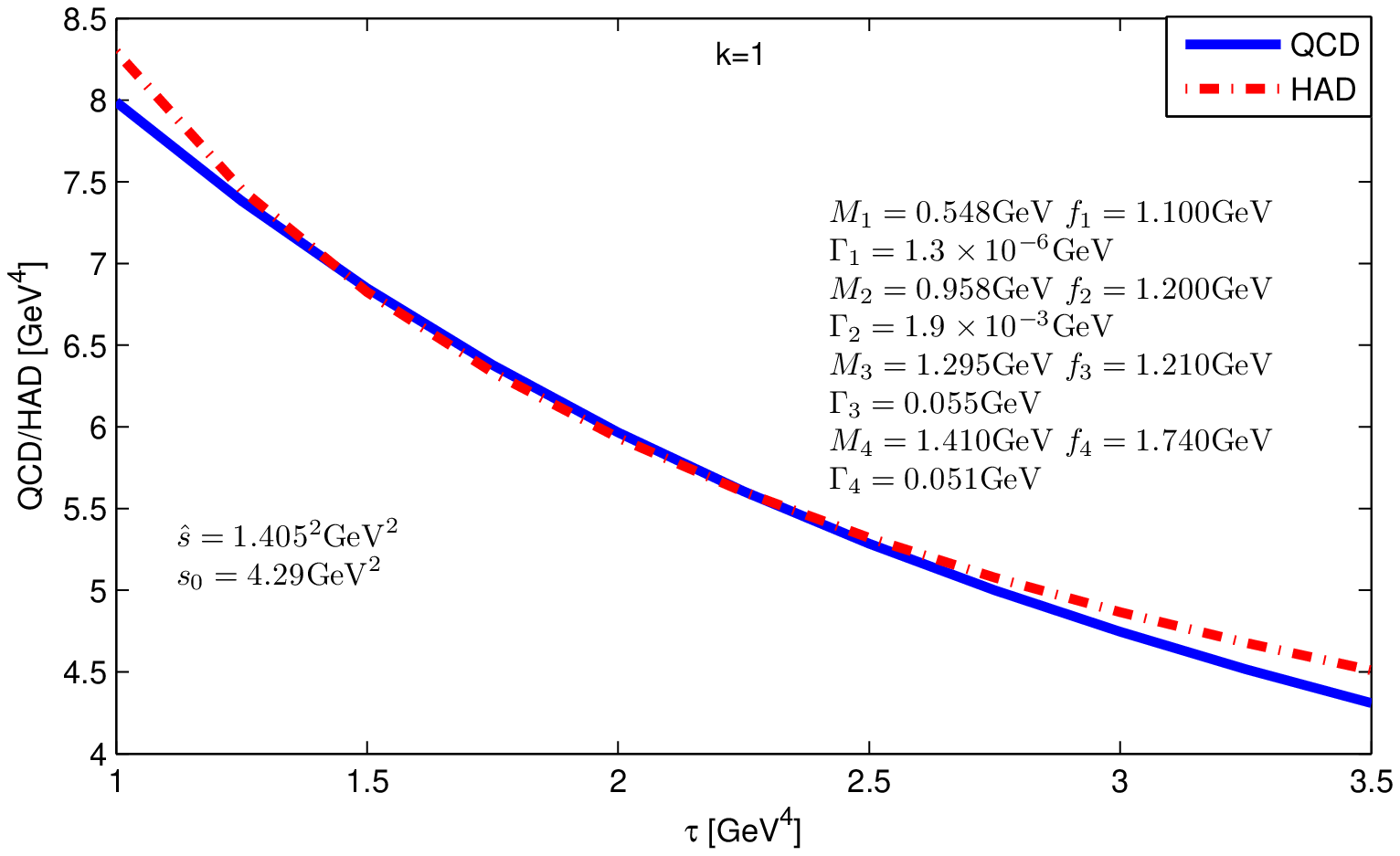}
\caption{\small The lhs (dashed line) and rhs (solid line) of the sum rules \eqref{eq4:43} with $k=-1,0,1$ versus $\tau$ in the case where the correlation function $\Pi^{\textrm{QCD}}(Q^2)$ contains the pure instanton, interference, and pure perturbative contributions, and the four finite-width resonance plus continuum model is adopted for the spectral function.}
\label{Fig:10}
\end{figure}

The optimal parameters of the four finite-width resonance model for the spectral function are listed in Tab \ref{tab:4BWR}, and the corresponding plots are shown in FIG. \ref{Fig:10}.
\begin{table*}[ht]
 \caption{\small {For the four finite-width resonances plus continuum model, the optimal fitting values of the mass $m$, width $\Gamma$, coupling constant $f$, continuum threshold $s_0$ , $\hat{s}$, and matching measure $\delta$ for the possible $0^{-+}$ resonances in the sum rule window $[t_{\textrm{min}},t_{\textrm{max}}]$ ($t=\tau^{-1}$) for the best matching between lhs and rhs of the sum rules with $k=-1,0,1$ are listed.}}
\begin{center}
\begin{tabular}{cccccccccc}\hline\hline
& $k$& resonances& $\sqrt{\hat{s}} (\textrm{GeV})$  &$m  (\textrm{GeV})$& $\Gamma  (\textrm{GeV})$& $f  (\textrm{GeV})$& $s_0 (\textrm{GeV}^2)$& $[t_{\textrm{min}},t_{\textrm{max}}](\textrm{GeV}^{-4})$& $\delta/10^{-4}$\\
\hline
     &   & $\eta(548)$&                &0.548& $1.3\times10^{-6}$& 1.100&       & \\
      &  $-1$& $\eta(958)$&  $1.405$   &0.958& $1.9\times10^{-3}$& 1.100&  5.30   & $[0.25,1.00]$& $2.3$\\
       &  & $\eta(1295)$&            &1.295&            0.055& 1.200&         &\\
       &   &               &           &1.412& 0.051& 1.340& &\\
\hline
      &    & $\eta(548)$&             &0.548& $1.3\times10^{-6}$& 1.200& & \\
      & $0$& $\eta(958)$& $1.500$  &0.958& $1.9\times10^{-3}$& 1.300& 4.80 & $[0.29,0.77]$& $2.2$\\
       && $\eta(1295)$ &           &1.295          & 0.055& 1.195&      &\\
       &   &            &             &1.420          & 0.051& 1.670&     &\\
\hline
     &    & $\eta(548)$&              &0.548& $1.3\times10^{-6}$& 1.100&   & \\
      & $1$& $\eta(958)$&   $1.405$ &0.958& $1.9\times10^{-3}$& 1.200& 4.29& $[0.42,0.83]$& $9.3$\\
       &   & $\eta(1295)$&            &1.295&            0.055& 1.210&     &\\
       &   &             &            &1.410&             0.051& 1.740&     &\\
\hline\hline \label{tab:4BWR}
\end{tabular}
\end{center}
\end{table*}

\bibliographystyle{unsrt}
\bibliography{Gaussian}

\end{document}